\documentclass[aps, preprint, nofootinbib,preprintnumbers,eqsecnum,superscriptaddress,sort,calc]
{revtex4}
\pdfoutput=1
\usepackage{color}

\usepackage[
      colorlinks=true,
      linkcolor=blue,
      urlcolor=blue,
      filecolor=black,
      citecolor=red,
      pdfstartview=FitV,
      pdftitle={},
        pdfauthor={Oscar Dias, Ramon Masachs},
        pdfsubject={},
        pdfkeywords={},
        pdfpagemode=None,
        bookmarksopen=true
      ]{hyperref}

\usepackage[normalem]{ulem}
\usepackage{amsmath}
\usepackage{enumerate}
\usepackage{amsfonts}
\usepackage{yfonts}

\usepackage{subfigure}
\usepackage{psfrag}

\usepackage{epsfig}
\usepackage[utf8]{inputenc}
\usepackage{float}
\usepackage{graphicx}
\usepackage{cancel}
\usepackage{mathrsfs}
\usepackage{amssymb}
\usepackage{amsfonts}
\usepackage{amsmath}
\usepackage{slashed}
\usepackage{comment}

\usepackage{graphicx}
\usepackage{bm}

\def\eg{{\it e.g. }}
\def\ie{{\it i.e. }}

\def\({\left(}
\def\){\right)}
\def\[{\left[}
\def\]{\right]}
\def\<{\langle}
\def\>{\rangle}
\newcommand{\be}{\begin{equation}}
\newcommand{\ee}{\end{equation}}
\newcommand{\bea}{\begin{eqnarray}}
\newcommand{\eea}{\end{eqnarray}}
\newcommand{\bwt}{\begin{widetext}}
\newcommand{\ewt}{\end{widetext}}

\newcommand{\bi}{\begin{itemize}}
\newcommand{\ei}{\end{itemize}}
\newcommand{\ben}{\begin{enumerate}}
\newcommand{\een}{\end{enumerate}}
\newcommand{\bca}{\begin{cases}}
\newcommand{\eca}{\end{cases}}
\newcommand{\bln}{\begin{align}}
\newcommand{\eln}{\end{align}}
\newcommand{\bst}{\begin{split}}
\newcommand{\est}{\end{split}}

\renewcommand{\Re}{\textrm{Re}\,}

\begin{document}

%\title {Superradiant hairy black holes in a Minkowski box}
%\title {Black  bombs in a Minkowski cavity and associated hairy black holes}
\title {Evading no-hair theorems: hairy black holes in a Minkowski box}
%\preprint{NSF-KITP }

\author{Oscar J. C. Dias}
\email{ojcd1r13@soton.ac.uk}
\affiliation{STAG research centre and Mathematical Sciences, University of Southampton, UK \vspace{1 cm}}

\author{Ramon Masachs}
\email{rmg1e15@soton.ac.uk}
\affiliation{STAG research centre and Mathematical Sciences, University of Southampton, UK \vspace{1 cm}}

\begin{abstract}
We find hairy black holes of Einstein-Maxwell theory with a complex scalar field that is confined inside a box in a Minkowski background. These regular hairy black holes are asymptotically flat and thus the presence of the box or mirror allows to evade well-known no-hair theorems. We also find the Israel surface stress tensor that the confining box must have to obey the energy conditions. In the zero horizon radius limit, these hairy black holes reduce to a regular asymptotically flat hairy soliton. We find our solutions using perturbation theory. At leading order, a hairy black hole can be seen as a Reissner-Nordstrom black hole placed on top of a hairy soliton with the same chemical potential (so that the system is in thermodynamic equilibrium). The hairy black holes merge with the Reissner-Nordstrom black hole family at the onset of the superradiant instability. When they co-exist, for a given energy and electric charge, hairy black holes have higher entropy than caged Reissner-Nordstrom black holes. Therefore, our hairy black holes are the natural candidates for the endpoint of charged superradiance in the Reissner-Nordstrom black hole mirror system.
\end{abstract}

\today

\maketitle

\tableofcontents

%\newpage

%\begin{figure}[ht]
%\centering
%\includegraphics[width=0.5\textwidth]{xxxxx.pdf}
%\caption{xxxxx}
%\label{fig:xxxxx}
%\end{figure}

%%%%%%%%%%%%%%%%%%%%%%%%%%%%%%%%%%%%%%%%%%%%%%%%%%%%%%%%%%%%%%%%%%%%%%%%%%%
\section{Introduction}\label{section:Introduction} %and summary of main results
%%%%%%%%%%%%%%%%%%%%%%%%%%%%%%%%%%%%%%%%%%%%%%%%%%%%%%%%%%%%%%%%%%%%%%%%%%%

In the 70's, Press and Teukolsky introduced the possibility of placing a Kerr black hole inside a mirror with reflecting boundary conditions \cite{Press:1972zz}. 
Bosonic fields scattering the central horizon can be amplified due to superradiance and then, at the mirror, reflected back to the central core of the spacetime. The multiple amplification/reflection process makes the system unstable. For this reason, Press and Teukolsky coined this system the ``black hole bomb" \cite{Press:1972zz}. 

If the system has electric charge, we can also have static black hole bomb systems  \cite{Herdeiro:2013pia,Hod:2013fvl,Degollado:2013bha,Hod:2014tqa,Li:2014gfg,Dolan:2015dha,Basu:2016srp,Ponglertsakul:2016wae,Ponglertsakul:2016anb,Sanchis-Gual:2015lje,Sanchis-Gual:2016tcm,Sanchis-Gual:2016ros,Hod:2016kpm,Fierro:2017fky,Li:2014xxa,Li:2014fna,Li:2015mqa,Li:2015bfa}. All we need is a complex scalar field with charge $q$ scattering a Reissner-Nordstr\"om black hole with chemical potential $\mu$ that is placed inside a box that reflects the scalar waves. Such a system is unstable to superradiance \cite{Herdeiro:2013pia,Hod:2013fvl,Degollado:2013bha,Hod:2014tqa,Li:2014gfg}. In addition it is also unstable to the near-horizon scalar condensation instability
\cite{Dias:2018zjg}.
The frequency spectrum and instability timescales of the Reissner-Nordstr\"om black hole bomb system were studied in detail in the literature \cite{Herdeiro:2013pia,Hod:2013fvl,Degollado:2013bha,Hod:2014tqa,Li:2014gfg,Hod:2016kpm,Dias:2018zjg}. The time evolution of the instabilities was also analysed  in recent studies \cite{Sanchis-Gual:2015lje,Sanchis-Gual:2016tcm,Sanchis-Gual:2016ros}. These Cauchy evolutions indicate that the original Reissner-Nordstr\"om-mirror system, when perturbed by a scalar field, evolves towards a charged hairy black hole with a scalar field floating above the horizon.  

In this manuscript we will study the  phase diagram of static asymptotically flat (regular) solutions of the Einstein-Maxwell theory with a complex scalar field confined inside a box. In particular, this can be the phase diagram in the microcanonical ensemble, where we plot the entropy of the solutions as a function of their mass and electric charge. To be meaningful, we use dimensionless quantities measured in units of the box radius $L$. A familiar member of this phase diagram is the Reissner-Nordstr\"om black hole placed inside the cavity. The interior solution of this caged Reissner-Nordstr\"om black hole was already discussed in the seminal works \cite{Gibbons:1976pt,Hawking:1982dh,Braden:1990hw} and, more recently, in \cite{Andrade:2015gja,Basu:2016srp}. This solution has vanishing scalar field but is linearly unstable to a scalar field perturbation, as discussed above. At the onset of the instability, the scalar field perturbation is regular {\it both} at the past and future event horizons of the black hole. Consequently, we might  expect that the back-reaction of this linear scalar perturbation to higher orders in perturbation theory results in a black hole solution that is regular everywhere and asymptotes to Minkowski spacetime. By continuity, this hairy black hole solution should extend away from the onset curve. That is to say, in the phase diagram, the instability onset curve of caged Reissner-Nordstr\"om black holes should also signal a bifurcation to a new branch of solutions of asymptotically flat hairy black holes. This new family should exist for a wide range of mass and electric charge. In particular, this should lead to non-uniqueness of solutions since hairy and caged Reissner-Nordstr\"om black holes should exist with the same mass and charge.

Given that the caged Reissner-Nordstr\"om black hole is unstable to the scalar field, we should further expect that, in the region of mass and charge where they co-exist, hairy black holes have higher (dimensionless) entropy than caged Reissner-Nordstr\"om  black holes. For if this is the case, an unstable caged Reissner-Nordstr\"om black hole would naturally evolve in time into the hairy black hole with the same mass and charge while preserving the second law of thermodynamics. That is to say, these hairy black holes should describe the endpoint of the time evolution simulations done in \cite{Sanchis-Gual:2015lje,Sanchis-Gual:2016tcm,Sanchis-Gual:2016ros}.

Additionally, it seems also reasonable to expect that the hairy black holes that we have been describing should have a zero horizon radius limit where they  reduce to a hairy soliton solution. This would be a horizonless asymptotically flat solution with a scalar field confined inside the Minkowski cavity. Again, gravitational collapse would be balanced by the electric repulsion. Under a $U(1)$ gauge transformation this hairy soliton has a dual boson star description whereby the scalar field with frequency $\omega$ has time dependence $e^{i \omega t}$. Such a solution is expected to exist since it should be the back-reaction of a normal mode that fits inside a Minkowski cavity (however the gravitoelectric fields would have no time dependence since they are sourced by the norm of a complex scalar field).  

To confirm these expectations we must solve the equations of Einstein-Maxwell theory with a complex scalar field that is forced to be confined inside the  box. It is the purpose of this manuscript to find the hairy black hole and solitonic solutions of this theory. Ref. \cite{Basu:2016srp} already discussed some thermodynamic properties of these hairy solutions in the grand-canonical ensemble. However, \cite{Basu:2016srp} constructed these solutions numerically and focused their attention on the description of the solutions only in the interior of the box. We will complement and extend their analysis in distinct directions. First, we find the hairy solutions analytically within perturbation theory which will allow to unveil physical aspects of the solution. More importantly, we will extend the solutions beyond the cavity and construct the full solution also in the exterior region. This is fundamental to find the energy-momentum of the box, \ie the Lanczos-Darmois-Israel  surface stress tensor that the cavity must have to yield a solution that obeys the Israel junction conditions and is thus continuous across the surface layer \cite{Israel:1966rt,Israel404,Kuchar:1968,Barrabes:1991ng}. To our best knowledge, this is the first instance where the Israel surface tensor is discussed in black hole bomb systems.

We will find that the discussion of black hole bombs is incomplete and even misleading without analysing this Israel surface stress tensor. For we will find that for general cavities, the associated Israel tensor does not obey the energy conditions, in particular the weak energy condition. In particular, to have a hairy black hole-mirror system that obeys the energy conditions we have to start with a box  in a Minkowski background that already has a specific energy-momentum content even before we place a soliton or horizon inside it. More concretely, the box needs to have a non-vanishing energy density and pressure. These quantities will be parametrized by a parameter $\eta$ that must have a non-vanishing value to yield a physical hairy black hole-mirror system. In view of our findings, it will be interesting to revisit the time evolutions of \cite{Sanchis-Gual:2015lje,Sanchis-Gual:2016tcm,Sanchis-Gual:2016ros} to discuss also the surface stress tensor of the system.

Refs. \cite{Dolan:2015dha,Ponglertsakul:2016wae,Ponglertsakul:2016anb} also discussed hairy black holes and solitons in the context of black hole bomb setups. However, their hairy solutions do not confine the scalar field inside the box. Consequently this scalar field can leak from the box and extend to the asymptotic region where it sources logarithmic divergences in the gravitoelectric fields (this is best discussed in Appendix A of our companion manuscript \cite{Dias:2018zjg}). Consequently, they are not asymptotically flat. Our solutions are fundamentally distinct since they are asymptotically flat (and regular everywhere). That is to say, our solutions are genuine asymptotically flat hairy solutions that evade the no-hair theorems of \cite{Ruffini:1971bza,Chrusciel:1994sn,Bekenstein:1996pn,Heusler:1998ua}. The presence of the box and its boundary conditions/surface stress tensor allow one to find that the Reissner-Nordstr\"om solution is not the only family of asymptotically flat, spherically symmetric and static (regular) black hole solutions of Einstein-Maxwell theory. 

An interesting byproduct of our analysis is that we will conclude that the thermodynamics and physics of black hole bomb systems is most appropriately discussed in terms of the Brown-York quasilocal thermodynamic quantities of the system  \cite{Brown:1992br}. Recall that the Israel surface stress tensor can be obtained integrating a Gauss-Codazzi equation along the normal direction to the surface layer \cite{Israel:1966rt,Israel404,Kuchar:1968,Barrabes:1991ng,MTW:1973}. Equivalently, it can be obtained taking the difference of the Brown-York quasilocal energy-momentum tensor outside and inside the surface layer  \cite{Brown:1992br}. These observations are the starting point to conclude that black hole bombs (and other systems with Israel boundary shells) must obey a quasilocal version of the first law of thermodynamics. This important property seems to have been missed in previous discussions of systems with Israel junction conditions. In particular, we propose that these gauge invariant quantities are relevant quantities to monitor in time evolution studies of black hole bomb systems.

The plan of this manuscript is as follows. In Section \ref{sec:Model} we describe our confining setup. Special attention is given to the discussion of the Israel junction conditions and Brown-York quasilocal charges. Hairy solitons are constructed within perturbation theory in Section \ref{sec:soliton}. The properties of Reissner-Nordstr\"om black holes confined in a box are discussed in Section \ref{sec:cagedRN}. Perturbation theory and a matched asymptotic expansion are  used to construct the hairy black holes in Section \ref{sec:HairyBH}. The physical properties of the solutions are then discussed in detail in Section \ref{sec:Physprop}. We describe  the thermodynamic properties of the solutions, the  Israel surface stress tensor of the system and analyse the necessary conditions to obey the energy conditions. Conclusions and final discussions are given in Section \ref{sec:conc}.

%%%%%%%%%%%%%%%%%%%%%%%%%%%%%%%%%%%%%%%%%%
\section{Einstein-Maxwell gravity with a confined scalar field}\label{sec:Model}

%%%%%%%%%%%%%%%%%%%%%
\subsection{Theory and setup}

We consider the action for Einstein-Maxwell gravity in four dimensions coupled to a charged scalar field:
\begin{align}\label{action}
S=\frac{1}{16 \pi G_N}\int{\mathrm d^4 x\sqrt{g}\left({\cal R}-\frac 1 2F_{\mu\nu}F^{\mu\nu}-2D_{\mu}\phi(D^{\mu}\phi)^{\dagger}+V(|\phi|)\right)},
\end{align}
where ${\cal R}$ is the Ricci scalar,  $A$ is the Maxwell gauge potential, $F=\mathrm d A$, and  $D_{\mu}=\nabla_{\mu}-i q A_{\mu}$ is the gauge covariant derivative of the system. We consider the potential $V(|\phi|)= m^2\phi\phi^{\dagger}$ with $m$ the mass of the scalar field. In our construction of hairy solutions, for concreteness we will take $m=0$, but our analysis could be straightforwardly extended to the massive case, $m>0$. We fix Newton's constant $G_N \equiv 1$. 

We are interested on solitonic and black hole solutions of \eqref{action} that are static, spherically symmetric and asymptotically flat. We can use reparametrizations of the time and radial coordinates, $ t\to \tilde{t}=t+H(t,r) $ and $r\to \tilde{r}(r)$, to fix the gauge to be such that the radius of a round $S^2$ is $r$ and there is no cross term $dt dr$ (this is known as the radial or Schwarzschild gauge). A field {\it ansatz} with the desired symmetries is
\begin{align}\label{fieldansatz}
\mathrm d s^2=-f(r)\mathrm d t^2+g(r)\mathrm d r^2+r^2\mathrm d \Omega_2^2, \qquad A_{\mu}\mathrm d x^{\mu}=A(r)\mathrm d t, \qquad \phi=\phi^{\dagger}=\phi(r),
\end{align} 
with $\mathrm d \Omega_2^2$ being the metric for the unit 2-sphere (expressed in terms of the polar and azimuthal angles $x=\cos\theta$ and $\varphi$).  We choose to work with the static {\it ansatz} \eqref{fieldansatz} where the scalar field is real. Note however that, should we wish,  we could perform a $U(1)$ gauge transformation with gauge parameter $\chi=-\omega t/q$,
\begin{align}\label{eq:U1gauge}
\phi=|\phi|e^{i\varphi} \to |\phi|e^{i(\varphi +q \,\chi)} \,,\qquad A_t \to A_t +\nabla_t \chi,
\end{align} 
to rewrite the scalar field as $\phi=|\phi|e^{-i\omega t}$, in which case we would be in a frame where the scalar field oscillates in time with a frequency $\omega$.\footnote{However, since the energy-momentum tensor of the scalar field only depends on $\phi \phi^\dagger$ and $\partial \phi (\partial \phi)^\dagger$, in the new gauge the gravitational and Maxwell fields would still be invariant under the action of the Killing vector field $\partial_t$.} 

As explained in the introduction, we introduce a confining box for the scalar field in order to trigger the instabilities in the system. We place this box at radius $L$. However, the system has the scaling symmetry:
\begin{align}\label{1scalingsym}
\begin{split}
\{t,r,x,\varphi\}\to\{\lambda_1 t,\lambda_1 r,x,\varphi\} , &\qquad \{f,g,A,\varphi\}\to \{f,g,A,\varphi\}, \\
\{q,L, r_+,m\}&\to \left\{\frac{q}{\lambda_1},\lambda_1 L, \lambda_1 r_+,\frac{m}{\lambda_1}\right\}
\end{split}
\end{align}
which rescales the line element and the gauge field 1-form as $\mathrm{d}s^2\to \lambda_1^2 \mathrm{d}s^2$ and $A \mathrm{d}t \to \lambda_1 \,A \mathrm{d}t$ but leaves the equations of motion invariant. We can use this scaling symmetry to work with dimensionless coordinates and measure thermodynamic quantities in units of $L$ (effectively this sets $L\equiv 1$),
\begin{equation}\label{adimTR}
T=\frac t L\,,\qquad R=\frac r L\,; \qquad R_+=\frac{r_+}{L}\,, \qquad e=q L\,, \qquad m_\phi=m L \,.
\end{equation}
The box is now at $R=1$.

The equations of motion obtained from extremising the action \eqref{action}  with $m=0$ are:
\begin{subequations}\label{eom}
\begin{align}
&g'(R)+g(R)^2 \left(\frac{1}{R}-\frac{e^2 R A(R)^2 \phi (R)^2}{f(R)}\right)+g(R) \left(-\frac{R A'(R)^2}{2 f(R)}-R \phi '(R)^2-\frac{1}{R}\right)=0,\\
& f'(R)-\frac{f(R) \left(g(R)+R^2 \phi '(R)^2-1\right)}{R}+\frac{R}{2} \left(A'(R)^2-2 e^2 A(R)^2 g(R) \phi (R)^2\right)=0,\label{feom}\\
&A''(R)+A'(R) \left(-\frac{f'(R)}{2 f(R)}-\frac{g'(R)}{2 g(R)}+\frac{2}{R}\right)-2 e^2 A(R) g(R) \phi (R)^2=0,\label{Aeom}\\
&\phi ''(R)+\frac{\phi '(R)}{2}  \left(\frac{f'(R)}{f(R)}-\frac{g'(R)}{g(R)}+\frac{4}{R}\right)+\phi (R)\frac{e^2 A(R)^2 g(R) }{f(R)}=0.\label{phieom}
\end{align}
\end{subequations}
Notice that from \eqref{feom} we can express $g(R)$ in terms of the other fields as
\begin{align}\label{gequation}
g(R)=\frac{R \left(R A'(R)^2+2 f'(R)\right)+f(R) \left(2-2 R^2 \phi '(R)^2\right)}{2 \left(e^2 R^2 A(R)^2 \phi (R)^2+f(R)\right)}.
\end{align}
We can substitute this expression into \eqref{eom}  to have a set of three differential equations that we solve for the fields $f(R),\ A(R)$ and $\phi(R)$. We can then use \eqref{gequation} to obtain straightforwardly $g(R)$.

In order to have a well-posed boundary value problem we must specify the boundary conditions in the inner and asymptotic boundaries of our spacetime. Additionally, we need to impose junction conditions at the timelike hypersurface $\Sigma$ at $R=1$ where the box is located. We are interested in asymptotically flat solutions with vanishing  scalar field at and outside this box, $\phi(R\geq 1)=0$.

The inner boundary can be the origin $R=0$ of our coordinate system (for horizonless solitonic solutions) or a Killing horizon with radius $R=R_+$ defined as the locus $f(R_+)=0$, when we have a black hole solution. The system is described by three second order ODEs which means that there are six arbitrary integration constants  when we do a Taylor expansion around the inner boundary. However, as a Dirichlet boundary condition we need to set three of these to zero in order to eliminate terms that would diverge at this boundary \cite{Dias:2015nua}. We are thus left with only three constants $f_0,A_0,\phi_0$ (say) such that the regular fields have the Taylor expansion:
\begin{equation}\label{BCorigin}
f(0)=f_0+\mathcal O(R),\qquad A(0)=A_0+\mathcal O(R),\qquad  \phi(0)=\phi_0+\mathcal O(R),
\end{equation}
at the origin of the soliton, or 
\begin{align}\label{BChorizon}
\begin{split}
& f(R_+)=f_0(R-R_+)+\mathcal O((R-R_+)^2),\\
& A(R_+)=A_0(R-R_+)+\mathcal O((R-R_+)^2),\qquad \phi(R_+)=\phi_0+\mathcal O((R-R_+)^2),
\end{split}
\end{align}
at the horizon radius of a black hole solution. Here we made the choice of working in the gauge where $A(R_+)=0$.
% (the chemical potential $\mu$ of the black hole is then simply $\mu=A |_{R\to\infty}\, ${\bf XXXXX}).

Consider now the outer boundary of our spacetime domain, namely the asymptotic infinity. Outside the box $\phi=0$ and the equations of motion are solved by the solution: $f^{out}(R)=c_f-\frac{M_0}{R}+\frac{\rho^2}{2 R^2},$ $A^{out}(R)=c_A+\frac{\rho}{R}$ and $g^{out}(R)=c_f/f^{out}(R)$ (onwards we use the superscript $^{out}$ to represent fields outside the box). Here, $c_f,M_0,c_A$ and $\rho$ are arbitrary integration constants. These are not constrained, \ie for any value of these constants we have an asymptotically flat solution.
However, the theory has a second scaling symmetry,
\begin{align}\label{2scalingsym}
\{T,R,x,\varphi\}\to\{\lambda_2 T, R,x,\varphi\} , \quad \{f,g,A,\varphi\}\to \{\lambda_2^{-2} f,g,\lambda_2^{-1} A,\varphi\}, \quad \{e, R_+\}\to \{e, R_+\},
\end{align}
that we can use to set $c_f=1$ so that $g$ approaches $1/f$ at large $R$. %The constant $c_A$ will be determined by fixing a specific condition for the gauge field at the box. This will be seen clearly later in the discussion around \eqref{eq:boxgaugecond} and when we discuss the soliton construction in \ref{sec:soliton}. $\rho$ is proportional to the electric charge of the solution.
Outside the box the solution to the equations of motion is then 
\begin{equation}\label{BCinfinity}
f(R)\big|_{R\geq 1}=1-\frac{M_0}{R}+\frac{\rho^2}{2 R^2}\,,\qquad A(R)\big|_{R\geq 1}=c_A+\frac{\rho}{R}\,,\qquad \phi(R)\big|_{R\geq1}=0\,,
\end{equation}
That is to say, outside the box  we require that the scalar field vanishes. Birkhoff's theorem for the Einstein-Maxwell theory then  states that the only static spherically symmetric asymptotic flat solution is described by the Reissner-Nordstr\"om solution \eqref{BCinfinity}. Note however that \eqref{BCinfinity} leaves three free integration constants, $M_0,c_A,\rho$, which will be determined once we have the solution inside the box. 

Our solutions are asymptotically flat. Therefore, some of the parameters  in \eqref{BCinfinity} are related to the ADM  conserved charges  \cite{Arnowitt:1962hi}.
Namely, the  adimensional ADM mass and electric charge of the system are given by (in units $G_N\equiv 1$):
\begin{align}\label{eq:thermogeneral}
\begin{split}
&M/L=\lim_{R\to \infty}\frac{R^2 f'(R)}{2\sqrt{f(R)  g(R)}}=\frac{M_0}{2},\\
&Q/L=\lim_{R\to \infty}\frac{R^2 A'(R)}{2 f(R) g(R)}=-\frac{\rho}{2}.
\end{split}
\end{align}
These ADM conserved charges measured at the asymptotic boundary include the contribution from the energy-momentum content of the box that confines the scalar hair. In the next subsection we discuss the effect of this box in more detail.

%%%%%%%%%%%%%
\subsection{Junction conditions at the box surface layer and associated Israel surface tensor}\label{subsec:JCisrael}

Our discussion of the boundary conditions at the inner and outer boundaries of our integration domain is complete. However, the requirement that the scalar field vanishes at the box $\Sigma$ located at $R=1$ ({\it and} outside it) comes with a cost. We will construct our hairy solutions perturbatively in the amplitude $\varepsilon$ of the scalar field. In our perturbation scheme, we define unambiguously the expansion parameter $\varepsilon$ to be such that $\phi^{out}(R\geq 1)=0$ and, in the interior of the box, the first derivative of the scalar field is $\varepsilon$ at {\it all} orders of the expansion,\footnote{Another natural choice to fix the perturbation scheme would have been to fix the expansion parameter to be the value of the scalar field at the horizon (black hole case) or at the origin (soliton case) at all orders. However, with this choice it would not be so straightforward to show that the zero horizon radius limit of the hairy black hole is the hairy soliton.}
\begin{equation}\label{def:epsilon} 
\phi^{in} \big|_{R=1}=\phi^{out} \big|_{R=1}=0,\qquad \phi^{out}(R)=0, \qquad \phi^{\prime\:in} \big|_{R=1}\equiv \varepsilon, 
\end{equation}
\ie for $R\leq 1$ the scalar field will be forced to have a Taylor expansion of the form $\phi\big|_{R=1}=\varepsilon (R-1)+\mathcal{O}(\varepsilon(R-1)^2)+\mathcal{O}(\varepsilon^3(R-1)^2)$. We are forcing a jump in the derivative of the scalar field normal to the cavity hypersurface. That is to say, the surface layer  $\Sigma$   must have a scalar charge density $4\pi \rho_\phi=-\partial_R \Phi \big |_{R=1}$ that sources this jump. The difference between the scalar field charge contained on a sphere just outside (where it vanishes) and just inside the surface layer is then  
\begin{equation}\label{surfScalarCharge} 
[Q_\phi]\equiv Q_\phi\big |_{R=1}^{out}-Q_\phi\big |_{R=1}^{in}=-\varepsilon
\end{equation}
which defines unambiguously $\varepsilon$.

Naturally, this forced condition on the scalar field has a further cost: we need to impose  junction conditions at the timelike hypersurface $\Sigma$  $-$ defined by $\frak{f}(R)=R-1=0$ and  with outward unit normal $n_{\mu}=\partial_\mu\frak{f}/|\partial \frak{f}|$ ($n_\mu n^\mu=1$) $-$ on the other fields of the system. Ideally, we would like to have a smooth crossing, \ie that the gravitational and gauge fields and their normal derivatives are continuous at $\Sigma$. But we are not guaranteed that this can be done.
To discuss this issue further it is good to set some notation.
From the perspective of an observer in the interior region, the cavity surface $\Sigma$ is parametrically described by $R=1$ and $T=T^{in}(\tau)=\tau$ such that the induced line element and induced gauge 1-form of the shell read
\begin{eqnarray}\label{SigmaIn}
{\mathrm d}s^2|_{\Sigma^{in}}&=&h_{ab}^{in} \,{\mathrm d}\xi^a {\mathrm d}\xi^b =- f^{in}|_{R=1}{\mathrm d}\tau^2+{\mathrm d}\Omega^2_2\,,\nonumber \\
 A |_{\Sigma^{in}}&=& a_{a}^{in} {\mathrm d}\xi^a = A^{in}|_{R=1}{\mathrm d}\tau\,,
\end{eqnarray}
where $\xi^a$ describe coordinates in $\Sigma$, $h_{ab}^{in}$ is the induced metric in $\Sigma$ and $a_a^{in}$ is the induced gauge potential in $\Sigma$. 
On the other hand, as seen from outside the cavity shell, $\Sigma$ is parametrically described by $R=1$ and $T=T^{out}(\tau)=N \tau$ so that the induced line element and the induced gauge 1-form are
\begin{eqnarray}\label{SigmaOut}
{\mathrm d}s^2 |_{\Sigma^{out}}&=&h_{ab}^{out} \,{\mathrm d}\xi^a {\mathrm d}\xi^b =- N^2 f^{out}|_{R=1}{\mathrm d}\tau^2+{\mathrm d}\Omega^2_2\,,\nonumber \\
 A |_{\Sigma^{out}}&=& a_{a}^{out} {\mathrm d}\xi^a = N A^{out}|_{R=1}{\mathrm d}\tau\,,
\end{eqnarray}
The reparametrization freedom parameter $N$ will be chosen as follows. We use the scaling symmetry \eqref{2scalingsym} in each region to set 
\begin{equation}\label{cf0}
f^{in}\big|_{R=1}=1\,,  \qquad f^{out}\big|_{R\to \infty}=1\,, 
\end{equation}
at all orders in $\varepsilon$; note that the second condition repeats the statement just before \eqref{BCinfinity}. An appropriate choice of the reparametrization freedom parameter $N$ will allow \eqref{cf0} to be obeyed.\footnote{Note that, in an alternative but equivalent scheme, we could choose to parametrize $\Sigma$ by $R=1$ and $T=T^{out}(\tau)=\tau$ and then transfer the reparametrization freedom encoded by $N$ into $f^{out}_0$ and $A^{out}_0$. In this case one would have 
$f^{out}\big|_{R\to \infty}\neq 1$ instead of \eqref{BCinfinity}, \eqref{cf} or \eqref{cfBH}.}  

The junction conditions required to join smoothly two spacetimes at a timelike hypersurface $\Sigma$ were studied by Israel \cite{Israel:1966rt,Israel404,Kuchar:1968,Barrabes:1991ng} built on previous work of Lanczos and Darmois. Next, we review these conditions. A solution is smooth at $\Sigma$ if and only if: 1) the induced metric $h_{ab}$ and induced gauge potential $a_a$  are continuous (\ie ${\mathrm d}s^2|_{\Sigma^{in}}={\mathrm d}s^2|_{\Sigma^{out}}$ and $A |_{\Sigma^{in}}=A |_{\Sigma^{out}}$), and 2)  the extrinsic curvature $K_{ab}$ (essentially the normal derivative of the induced metric) and the normal derivative of the induced gauge field, $f_{aR}$, are continuous.  If we denote the solution inside (outside) $\Sigma$ by the superscript $^{in}$ ($^{out}$), the Israel junction conditions can be written as
\begin{subequations}\label{IsraelJunctionConditions}
\begin{align}
& a_{a}^{in}\big|_{R=1}=a_{a}^{out}\big|_{R=1}\,, \label{eq:Israeljunction1} \\
&h_{ab}^{in}\big|_{R=1}=h_{ab}^{out}\big|_{R=1}\,; \label{eq:Israeljunction2} \\
& f_{aR}^{in}\big|_{R=1}=f_{aR}^{out}\big|_{R=1}\,,\label{eq:Israeljunction3} \\
&K_{ab}^{in}\big|_{R=1}=K_{ab}^{out}\big|_{R=1}; \label{eq:Israeljunction4}
\end{align}
\end{subequations}
where $h_{ab}=g_{ab}-n_a n_b$ is the induced metric at $\Sigma$ and  $K_{ab}=h_a^{\phantom{a}c}\nabla_c n_b$ is the extrinsic curvature.

If, as it is our case, the exterior curvature condition \eqref{eq:Israeljunction4} is not satisfied then the solution is singular at $\Sigma$. This is interpreted as due to the presence  of a  Lanczos-Darmois-Israel surface stress tensor ${\cal S}_{ab}$ at the hypersurface layer proportional to the difference of the extrinsic curvature on both sides of the hypersurface. More concretely, the  Lanczos-Darmois-Israel surface stress tensor induced in $\Sigma$ is  \cite{Israel:1966rt,Israel404,Kuchar:1968,Barrabes:1991ng}
\begin{align}\label{eq:inducedT}
{\cal S}_{ab}=-\frac{1}{8 \pi}\Big([K_{ab}]-[K] h_{ab}\Big),
\end{align}
where $K$ is the trace of the extrinsic curvature and $[K_{ab}]\equiv K_{ab}^{out}\big|_{R=1}-K_{ab}^{in}\big|_{R=1}$. This surface tensor is the pull-back of the energy-momentum tensor integrated over a small region around the hypersurface $\Sigma$ \ie it is obtained by integrating the appropriate Gauss-Codazzi equation  \cite{Israel:1966rt,Israel404,Kuchar:1968,Barrabes:1991ng,MTW:1973}.
Essentially, \eqref{eq:inducedT} describes the energy-momentum tensor of the cavity (the ``internal structure" of the mirror) that we have to build to confine the scalar field inside. With our explicit construction of the hairy solutions of the system we will be able to compute this  Lanczos-Darmois-Israel stress tensor. To our best knowledge, this gives the first explicit description of the box matter content of a black hole bomb.

Lanczos-Darmois-Israel surface stress tensor obeys the conservation law  \cite{Israel:1966rt,Israel404,Kuchar:1968,Barrabes:1991ng},
\begin{align}\label{eq:conservationIsrael}
{\cal D}^b {\cal S}_{ab}=-T_{\mu \nu}h^{\mu}_{\phantom{\mu}a}n^\nu 
\end{align}
where ${\cal D}_a$ is the covariant derivative w.r.t. the 3-dimensional induced metric $h_{ab}$ on $\Sigma$, $h^{\mu}_{\phantom{\mu}a}$ is the 4-dimensional tensor that projects quantities onto the hypersurface $\Sigma$, and $T_{\mu\nu}$ the energy-momentum tensor of \eqref{action}.\footnote{For our hairy system, $n^{\mu} T_{\mu\nu}h^\nu_{\phantom{\nu} b}=0$. As such \eqref{eq:conservationIsrael} boils down to ${\cal D}^b {\cal S}_{ab}=0$.} This follows from the definition \eqref{eq:inducedT} and from the ADM Hamiltonian constraint (\ie from the contracted Gauss-Codazzi equation along the normal to $\Sigma$). On the other hand, the ADM momentum constraint, \ie the normal-normal contracted Gauss-Codazzi equation yields the relation \cite{Israel:1966rt,Israel404,Kuchar:1968,Barrabes:1991ng}
\begin{align}\label{eq:conservationIsrael}
\frac{1}{2}\left(K_{ab}^{out}+K_{ab}^{in}\right){\cal S}^{ab}=\left[ T_{\mu \nu}n^\mu n^\nu \right]
\end{align}
where the square brackets again represent the difference between the RHS quantity just outside and just inside $\Sigma$. 

In our case the  Lanczos-Darmois-Israel surface stress tensor \eqref{eq:inducedT} ensures that the scalar hair is confined to the interior of the box with radius $R=1$. The outward normal to $\Sigma$ is $n=\sqrt{g}\,dr$, the 3-dimensional induced metric on $\Sigma$ is $h_{ab}=g_{ab}$ (for $a,b=\{t,\theta,\phi\}$) and the non-vanishing extrinsic curvature components are
\begin{equation}\label{Kab}
K^{t}_{\phantom{t}t}=-\frac{f'(R)}{2f(R)\sqrt{g(R)}}\,,
 \qquad K^{i}_{\phantom{i}j}=\frac{1}{R\sqrt{g}}\,\delta^{i}_{\phantom{i}j}\,,\\
\end{equation} 
%K^{t}_{\phantom{t}t}=\frac{R^2 \left[2 f \left(\phi '\right)^2+2 q^2 A^2 g\, \phi ^2-\left(A'\right)^2\right]+2 f (g-1)}{4 R\,f \sqrt{g}},
with $\delta^{i}_{\phantom{i}j}$ being the Kronecker symbol ($i,j=\theta,\phi$). 
One of the main results of our study will be that we {\it cannot} make the extrinsic curvature \eqref{Kab} continuous across $\Sigma$ if we require that the scalar field vanishes at and outside the box $-$ see \eqref{def:epsilon} $-$ as is required for the black hole bomb system.

%%%%%%%%%%%%%
\subsection{Brown-York quasilocal formalism and the Israel surface stress tensor}\label{subsec:quasilocal}
The Lanczos-Darmois-Israel surface stress tensor \eqref{eq:inducedT} was originally derived by integrating the Gauss-Codazzi equations along the direction orthogonal to the thin surface layer \cite{Israel404,Israel:1966rt,MTW:1973}. However, it can be equivalently derived using the quasilocal Brown-York formalism \cite{Brown:1992br}.  This derivation further enlightens the physical interpretation of the tensor \cite{Brown:1992br}. Therefore, here we will highlight some of the key properties of the Brown-York energy-momentum tensor and its relation with the  Lanczos-Darmois-Israel tensor.

The Brown-York surface  energy-momentum stress tensor of a $R=const$ timelike hypersurface $\Sigma$ (with unit normal $n$, induced metric $h_{ab}$ and extrinsic curvature $K_{ab}$) $-$ \eg our cavity wall $-$ is given by \cite{Brown:1992br}
\begin{align}\label{BrownYork}
{\cal T}_{ab}=-\frac{1}{8 \pi}\Big(K_{ab}-K h_{ab}\Big).
\end{align}
It follows from the Einstein equation, $G_{\mu\nu}=8\pi T_{\mu\nu}$, that the surface energy-momentum tensor ${\cal T}_{ab}$ obeys the conservation law ${\cal D}^a {\cal T}_{ab}=n^{\mu} T_{\mu\nu}h^\nu_{\phantom{\nu} b}$, which in our case reduces simply to ${\cal D}^a {\cal T}_{ab}=0$ since $n^{\mu} T_{\mu\nu}h^\nu_{\phantom{\nu} b}=0$.\footnote{The  quantities ${\cal D}_a$ and $h^\nu_{\phantom{\nu} b}$ were already defined below  \eqref{eq:conservationIsrael}.  After \eqref{IsraelBY} it will be clear that \eqref{eq:inducedT} follows from \eqref{BrownYork}.}

The Brown-York energy, momentum and spatial stress surface densities on $\Sigma$ are computed introducing a $T=const$ spacelike hypersurface $\Sigma_T$ with unit normal $u$ ($u^2=-1$; in our case $u=\sqrt{f}dT$) that intersects orthogonally the timelike hypersurface $\Sigma$ at a 2-dimensional surface $B$ (in our case a 2-sphere of radius $R$, $B\equiv S^2$).
The statement that $\Sigma_T$ is orthogonal to $\Sigma$ means that $u\cdot n |_\Sigma=0$. Then the unit normal $n$ in spacetime to the three-boundary $\Sigma$ is also the unit normal in $\Sigma_T$ to the two-boundary $B$. In these conditions, let   $\sigma^{ab}=g^{ab}-n^a n^b +u^a u^b$ be the induced metric on $B$, $\sigma$ be the associated determinant, and let ${i,j}$ run over the coordinates of the 2-sphere such that $\sigma^a_{\phantom {a} i}$ is the 3-dimensional tensor that projects quantities onto $B$. Then the Brown-York surface energy density $\tilde{\rho}$, surface momentum density $\tilde{j}_i$ and spatial stress surface densities $\tilde{p}_{ij}$ are
\begin{subequations}\label{BYdensities0}
\begin{align}
& \tilde{\rho}\equiv u_a u_b {\cal T}^{ab}=- {\cal T}^{T}_{\phantom{T}T},\\
&  \tilde{j}_i\equiv -\sigma_{i a}u_b {\cal T}^{ab},\\
&  \tilde{p}_{ij}\equiv \sigma_{i a}\sigma_{j b} {\cal T}^{ab}.
\end{align}
\end{subequations}
To these expressions, we must still apply the appropriate reference background (denoted by the subscript $_0$) subtraction procedure detailed in \cite{Brown:1992br}. For example, the physical energy surface density is $ \rho=\tilde{\rho}-\tilde{\rho}_0$ and similarly for the other quantities. In our case the reference background for this subtraction procedure is, naturally, the Minkowski spacetime. Once this is done we find that, for our system, the Brown-York surface charge densities on a 2-sphere with radius $R$ are:
\begin{subequations}\label{BYdensities}
\begin{align}
& \rho L=\frac{1}{4\pi}\frac{1}{R}\left( 1-\frac{1}{\sqrt{g}}\right),\\
& j_i=0,\\
& p^i_{\phantom{i}j}/L= \frac{R^2}{8\pi}\left( \frac{f'}{2f\sqrt{g}}+\frac{1}{R\sqrt{g}}-\frac{1}{R}\right)\sigma^i_{\phantom{i}j}\,.
\end{align}
\end{subequations}

The Brown-York quasilocal mass contained inside a 2-sphere with radius $R=1$ is then ($G_N\equiv 1$)
\begin{eqnarray}\label{BYmass}
{\cal M}/L&=&\int_B d^2x \sqrt{\sigma} \rho \nonumber\\
  &=&  R\left( 1-\frac{1}{\sqrt{g}}\right)\Big|_{R=1}.
\end{eqnarray}
Note that, as required for a good definition of quasilocal mass, when we send the $S^2$ radius to infinity, $R\to\infty$, the Bown-York mass \eqref{BYmass} reduces to the ADM mass \eqref{eq:thermogeneral}. This is explicitly checked using \eqref{gequation} and \eqref{BCinfinity}.

As highlighted in \cite{Brown:1992br}, by construction, the Lanczos-Darmois-Israel surface energy tensor ${\cal S}_{ab}$ \eqref{eq:inducedT} of the thin shell $\Sigma$ (with $R=1$ in our case) is given by the difference between the Brown-York surface tensor just outside and inside the surface layer
\begin{eqnarray}\label{IsraelBY}
{\cal S}_{ab}\equiv \left[ {\cal T}_{ab}\right]={\cal T}_{ab}\big |_{R=1}^{out} -{\cal T}_{ab}\big |_{R=1}^{in}\,,
\end{eqnarray}
That is to say, the Lanczos-Darmois-Israel surface stress tensor gives the stress energy-momentum tensor of the surface layer. 

The Brown-York quasilocal charge inside a 2-sphere with radius $R=1$ follows from Gauss's law evaluated at the spherical boundary,
\begin{eqnarray}\label{BYcharge}
{\cal Q}/L|_{R\to1}  &=&\frac{1}{8\pi}\int_{\Sigma}\star F  \nonumber\\
  &=& \sqrt{g(R)}\frac{R^2A'(R)}{2 g(R)f(R)}\Big|_{R=1}.
\end{eqnarray}

To complete the thermodynamic description of our solutions we still need to define the chemical potential, temperature and entropy.
The chemical potential is defined as the difference between the gauge potential at the box and at the horizon, 
\begin{align}\label{eq:QLchp}
\mu=A(1)-A(R_+).
\end{align}

Finally the temperature and the entropy are defined from the surface gravity at the horizon and from the  horizon area:

\begin{align}
T_H L=\lim_{R\to R_+}\frac{f'(R)}{4\pi\sqrt{f(R) g(R)}},\qquad S/L^2=\pi R_+^2.
\end{align}

The quasilocal mass and electric charge must satisfy a quasilocal form of the first law of thermodynamics: 
\begin{align}
&\mathrm d {\cal M}=T_H \,\mathrm d S+\mu \, \mathrm d {\cal Q}, \quad \text{for black holes},\label{1stlawBH}\\
&\mathrm d {\cal M}=\mu \,\mathrm d {\cal Q}, \qquad \qquad \ \ \text{for solitons.}\label{1stlawsoliton}
\end{align}
We will use these relations as a non-trivial check of our solutions.

%%%%%%%%%%%%%%%%%%%%%%%%%%
\section{Small solitons (boson stars) confined in a box}\label{sec:soliton}

Consider a box $\Sigma$ placed at $r=L$ (\ie $R=1$) in a Minkowski background with a constant gauge field $A=A_0 \,\mathrm{d}T$. We can now consider adding a scalar field  $\phi(R)$  to get an asymptotically flat hairy soliton that is regular at the origin and vanishes at and outside the box, $\phi(R\geq 1)=0$. Here, we will construct this solution perturbatively in the amplitude $\varepsilon$ of the scalar field. In our pertubation scheme, we choose unambiguously our expansion parameter to be $\varepsilon\equiv \phi'(R=1)$ at {\it all} expansion orders in $\varepsilon$ (\ie there are no corrections to  $\phi'(R=1)$ at order $\varepsilon^2$ or higher). Also, we find the energy-momentum content, \ie the Israel-Darmois surface stress tensor, that the thin shell $\Sigma$ must have to yield a physical setup that obeys the energy conditions.  

Not all perturbations fit inside the box. Those that do so, \ie the normal modes of the system, have their dimensionless frequency $\Omega_{p}\equiv \omega_p L$ quantized as
\begin{align}
\Omega_{\ell,p}=\sqrt{j_{\ell+\frac{1}{2},p}^2+m_{\phi}^2}-A_0 e
\end{align}
as described in \cite{Dias:2018zjg}. Here, $m_\phi$ and $e$ are the dimensionless mass and charge of the scalar field, respectively, and  $\ell$ and $p$ are the angular and radial quantum numbers that give the number of nodes of the normal mode along the polar and radial directions, respectively. Finally, $ j_{\ell+\frac{1}{2},p}$ is the location of the zeros of the Bessel function $J_\nu(z)$,  for $\nu=\ell+\frac{1}{2} \in \mathbb{R}$.  We work in a gauge where the scalar field is static, \ie its quantized frequency vanishes at the expense of fixing an appropriate chemical potential $A_0$ for the background field. In these conditions, the ground state solution (\ie with lowest  energy, $p=1$)  for a spherically symmetric ($\ell=0$) massless scalar field ($m_\phi=0$) is  described by $\Omega_{0,1}=j_{\frac{1}{2},1}-A_0 e=\pi-A_0 e$. In the static gauge, the background gauge field is  given  
by $A_0=\frac{\pi}{e}$. This is the case we will describe in detail. Solutions with different parameters $\ell, p, m_\phi$ can be constructed in a similar way.

We can now ask whether we can back-react this normal mode solution to higher orders in perturbation theory (\ie to higher orders in the amplitude $\varepsilon$) and eventually at full non-linear order. This question has a positive answer if we can keep the solution with the desired asymptotics regular at the origin at all orders. If so, the theory admits a soliton solution. Naturally, as the order of the expansion grows beyond linear order, the non-linearities of the field equations imply that the scalar field sources corrections on the gravitational and electric field. Therefore, not only $\phi$ but also $f$ and $A$ have an expansion in the amplitude $\varepsilon$. Furthermore, note that although the scalar field vanishes outside the box at all perturbation orders, the gravitational and electric fields outside the box will nevertheless be corrected at each order as a consequence of requiring the induced fields to be continuous at the box. In addition, these Israel junction conditions will determine the energy-momentum tensor that the box at $R=1$ must have to be able to accommodate such a solution. The purpose of this Section is to  construct perturbatively this soliton.    

In the conditions just described, the fields of the soliton solution have the expansion
\begin{align}\label{eq:schemeSoliton}
\begin{split}
& f^{(\frak{R})}(R)=\sum_{n\geq 0}\varepsilon^{2n}f^{(\frak{R})}_{2n}(R),  \\
&A^{(\frak{R})}(R)=\sum_{n\geq 0}\varepsilon^{2n}A^{(\frak{R})}_{2n}(R), \\
& \phi^{(\frak{R})}(R)=\sum_{n\geq 0}\varepsilon^{2n+1}\phi^{(\frak{R})}_{2 n+1}(R), \\
\end{split}
\end{align}
where the superscript $^{(\frak{R})}$ indicates whether we are considering the region inside ($^{(\frak{R})=in}$) or outside ($^{(\frak{R})=out}$) the box located at $R=1$. We use the scaling symmetry \eqref{2scalingsym} in each region to impose \eqref{cf0} at all orders in $\varepsilon$, \ie
\begin{equation}\label{cf}
f^{in}\big|_{R=1}=1\,,  \qquad f^{out}\big|_{R\to \infty}=1\,.
\end{equation}

In the interior region, as described in \eqref{SigmaIn}, the cavity surface $\Sigma$ is parametrically described by $R=1$ and $T=T^{in}(\tau)=\tau$ such that the induced line element of the shell reads ${\mathrm d}s^2|_{\Sigma^{in}}=- f^{in}|_{(R=1)}{\mathrm d}\tau^2+{\mathrm d}\Omega^2_2$ and the induced gauge 1-form is $A |_{\Sigma^{in}}= A^{in}|_{(R=1)}{\mathrm d}\tau$. 
At leading order ($n=0$) we have a scalar field perturbation around the Minkowski background ($f^{in}_0=1$) with a constant gauge potential $A^{in}_0=a_0$. 
The most general solution of the associated Klein-Gordon equation is $\phi(R)=R^{-1}\left(\beta_1 e^{-i e a_0 R}+\beta_2 e^{i e a_0  R}\right)$. To avoid a divergence $1/R$ at the origin we must choose $\beta_2=-\beta_1$. At the box, the condition $\phi(R=1)=0$ $-$ see \eqref{def:epsilon} $-$ requires $A^{in}_0=a_0=p\frac{\pi}{e}$ for $p=1,2,3,\cdots$. Essentially, $p$ dictates the number of radial nodes of the soliton we are interested in finding. For concreteness, onwards we focus on the $p=1$ case. This will give the ground state soliton with lowest energy, i.e. the soliton that corresponds to the back-reaction of the spherically symmetric ($\ell=0$) normal mode of AdS with the lowest frequency.   Finally, as discussed in \eqref{def:epsilon}, we define unambiguously our expansion parameter $\varepsilon$ to be such that $\phi(R=1)=\varepsilon(R-1)$ (at {\it all} orders in perturbation theory). This requires that $\beta_1=-i/(2\pi)$. In these conditions, the regular fields inside the box read
\begin{equation}\label{solitonorder0}
f^{in}_0=1,\qquad A^{in}_0=\frac{\pi}{e},\qquad \phi^{in}_1(R)=-\frac{\sin(\pi R)}{\pi R}\,.
\end{equation}

As discussed in \eqref{SigmaOut}, when seen from outside, the cavity shell $\Sigma$ is parametrically described by $R=1$ and $T=T^{out}(\tau)=N \tau$ so that the induced line element is ${\mathrm d}s^2 |_{\Sigma^{out}}=- N^2 f^{out}|_{(R=1)}{\mathrm d}\tau^2+{\mathrm d}\Omega^2_2$ and the induced gauge 1-form is $A |_{\Sigma^{out}}=N A^{out}|_{(R=1)}{\mathrm d}\tau$.
Solving the equations of motion outside the box at leading order yields:
\begin{equation}
f^{out}_0(R)=C_2^{f_0}-\frac{\eta}{R}+\frac{(C_1^{A_0})^2}{2 R^2},\qquad A^{out}_0(R)=\frac{C_1^{A_0}}{R}+C_2^{A_0}.
\end{equation}
The boundary conditions \eqref{BCinfinity}, \ie \eqref{cf}, imply that $C_2^{f_0}=1$. On the other hand the junction conditions  \eqref{eq:Israeljunction1}-\eqref{eq:Israeljunction3}  at the box,
\begin{align}
\begin{split}
&f^{in}_0(1) \,{\mathrm d}\tau^2=N^2 f^{out}_0(1)\,{\mathrm d}\tau^2 \qquad \Leftrightarrow \quad 1=N^2\left(1-\eta+\frac{(C_1^{A_0})^2}{2}\right),\\
&A^{in}_0(1) \,{\mathrm d}\tau=N A^{out}_0(1)\,{\mathrm d}\tau 
\qquad \quad\Leftrightarrow \quad  \frac{\pi}{e} =N \left(C_2^{A_0}-C_1^{A_0}\right),\\
& A^{in\, \prime}_0(1) \, {\mathrm d}\tau {\mathrm d}R= N A^{out\, \prime}_0(1) \,{\mathrm d}\tau {\mathrm d}R  \quad \Leftrightarrow \quad 0=C_1^{A_0} N\,.
\end{split}
\end{align}
fix three other integration constants (including the reparametrization factor):
\begin{align}
C_1^{A_0}=0,\qquad C_2^{A_0}=\frac{\pi}{e}\frac{1}{N},\qquad N=\frac{1}{\sqrt{1-\eta}}\,.
\end{align}
The integration constant $\eta$ is undetermined. It characterizes the energy-momentum content of the box as discussed below.
Altogether the solution at leading order ($n=0$) is
\begin{align}\label{eq:solitonleadeingterms}
\begin{split}
&f^{in}_0(R)=1\,,\qquad \qquad \quad A^{in}_0(R)=\frac{\pi}{e}\,,\qquad \qquad \qquad\phi^{in}_1(R)=-\frac{\sin(\pi R)}{\pi R}\,; \\
& f^{out}_0(R)=1-\frac{\eta}{R}\,, \qquad  A^{out}_0(R)=\frac{\pi}{e}\sqrt{1-\eta}\,, \qquad \phi^{out}_1(R)=0\,;\\
&N=\frac{1}{\sqrt{1-\eta}}\,.
\end{split}
\end{align}

In these conditions, the leading order $n=0$ Lanczos-Darmois-Israel surface stress tensor \eqref{eq:inducedT} has the non-vanishing components
\begin{equation}\label{IsraelS0}
 \mathcal{S}^t_{\phantom{t}t}=-\frac{1}{4\pi}\left( 1-\sqrt{1-\eta}\right)+\mathcal{O}(\varepsilon^2)\,,\qquad \mathcal{S}^i_{\phantom{i}i}=\left(\frac{1-\frac{1}{2}\eta}{8\pi\sqrt{1-\eta}}-1\right)+\mathcal{O}(\varepsilon^2)\,,
\end{equation} 
where $i=2,3$ runs over the $S^2$ coordinates, and $\mathcal{O}(\varepsilon^2)$ reminds us that this tensor will receive corrections at order $n=1$.
The reader will observe that we have not imposed the final Israel junction condition \eqref{eq:Israeljunction4}. We {\it could} do it and thus fix the leftover integration constant $\eta$ to vanish to have a continuous extrinsic curvature across $\Sigma$ and thus a vanishing Lanczos-Darmois-Israel tensor \eqref{IsraelS0} at zeroth order in the expansion. Had we chosen to do so, at leading order the surface layer $\Sigma$ would have no energy density neither pressure. Thus, it would just split the spacetime into interior and exterior regions that are both described by the Minkowski solution (with a constant gauge field) with the interior region also containing a linear scalar field that, at order $n=0$, has not yet back-reacted on the gravitoelectric fields neither on the box surface stress tensor. However, we will {\it not} fix $\eta$, \ie we will not require the extrinsic curvature to be continuous at leading order $n=0$. The reason being that at order $n=1$ we will find that the Lanczos-Darmois-Israel surface stress tensor gets negative contributions proportional to $\varepsilon^2$ that would violate all the energy conditions! (Indeed see the final result \eqref{IsraelSoliton}.) So to have a {\it physical} box that confines the scalar field inside it in the conditions \eqref{def:epsilon}, we must choose a box that at leading order $n=0$ has a non-vanishing Lanczos-Darmois-Israel surface tensor \eqref{IsraelS0}. The associated positive energy density and pressure will be able to accommodate the higher order negative contributions and allow the energy conditions to be obeyed. In these conditions, if $0<\eta\leq 1$ the ADM mass is non-vanishing: the surface layer has energy-momentum surface tensor described by \eqref{IsraelS0} (\ie $\eta$) which has an associated mass that can be read asymptotically. Therefore, the interior solution is Minkowski spacetime with a linear scalar field and the exterior solution is described by the Schwarzschild geometry with ADM mass proportional to $\eta$.\footnote{Actually, apart from the presence of scalar field that  motivates here the need for an Israel surface layer, this setup is the textbook example of a Lanczos-Darmois-Israel  static thin shell separating Minkowski spacetime in the interior from the Schwarzschild geometry in the exterior: see \eg section 3.10 of Poisson's textbook  (with rotation $a=0$) \cite{Poisson:2004}.} The hairy solitons are thus a 2-parameter family of solutions described by the expansion parameter $\varepsilon$ and by the energy density of the cavity (that also fixes uniquely its pressure). Note that we are choosing the simplest physical box that can confine the scalar field: it has no surface electric charge density. If we wish, this cavity can also have an electric charge density  in which case the exterior solution would be described by the Reissner-Nordstr\"om geometry (\ie the hairy solitons would be a 3-parameter family of solutions).

This completes the analysis up to $\mathcal O(\varepsilon^1)$. 
These fields source the equations of motion at next orders. The following non-trivial equations for the fields are at order $\mathcal O(\varepsilon^2)$ for the gravitational and gauge fields and $\mathcal O(\varepsilon^3)$ for the scalar field, \ie $n=1$. 

At $\mathcal O(\varepsilon^2)$, in the box interior $R\leq  1$ one has:
\begin{align}
\begin{split}
&f^{in}_2(R)=\frac{B_1^{f_2}}{R}+B_2^{f_2}+2\left[\,\ln R- \text{Ci}(2 \pi  R)\right]+\frac{\sin (2 \pi  R)}{\pi  R},\\
&A^{in}_2(R)=\frac{B_1^{A_2}}{R}+B_2^{A_2}+\frac{e}{\pi}\left[\,\ln R-\text{Ci}(2 \pi  R)\right]+\frac{e}{2\pi^2} \frac{\sin (2\pi  R)}{R},
\end{split}
\end{align}
where $\text{Ci}(x)=-\int_x^{\infty}\frac{\cos z}{z}\mathrm d z$ is the cosine integral function  and $B_{1,2}^{f_2}$, $B_{1,2}^{A_2}$ are integration constants. Regularity at $R=0$ implies that $B_1^{f_2}=0$ and $B_1^{A_2}=0$. On the other hand, the first condition in \eqref{cf} fixes $B_2^{f_2}=2\text{Ci}(2 \pi)$.

Moving to the exterior region, at  $\mathcal O(\varepsilon^2)$ the most general solution is described by:
\begin{equation}
f^{out}_2(R)=C_2^{f_2}-\frac{C_1^{f_2}}{R},\qquad A^{out}_2(R)=C_2^{A_2}-\frac{C_1^{A_2}}{R}.
\end{equation}
where $C_{1,2}^{f_2}$ and $C_{1,2}^{A_2}$ are integration constants.
The boundary conditions \eqref{BCinfinity} fix $C_2^{f_2}=0$.
%\begin{align}
%\begin{split}
%&h_{ab}^{in}\big|_{R=1}{\mathrm d}\xi^a {\mathrm d}\xi^b=h_{ab}^{out}\big|_{R=1}{\mathrm d}\xi^a {\mathrm d}\xi^b,\\
%&a^{in}_a {\mathrm d}\xi^a=a^{out}_a {\mathrm d}\xi^a \,,\\
%& f_{aR}^{in}{\mathrm d}\xi^a =f_{aR}^{out}{\mathrm d}\xi^a \,.
%\end{split}
%\end{align}

To impose the junction conditions  we first do a Taylor expansion of the fields at the box:
\begin{align}
\begin{split}
&f^{in}_2(1)=0\,,\\
&N^2 f^{out}_2(1)=-\frac{C_1^{f_2}}{\sqrt{1-\eta}}\,;\\
&A^{in}_2(R)|_{R\to 1}= B_2^{A_2}-\frac{e}{\pi} \text{Ci}(2 \pi )+\frac{e}{\pi}(R-1)+\mathcal O\left((R-1)^2\right),\\
&N  A^{out}_2(R)|_{R\to 1}=  B_2^{A_2}-B_1^{A_2} 
+B_1^{A_2} (R-1)+\mathcal O\left((R-1)^2\right)\,.
\end{split}
\end{align} 
The junction conditions \eqref{eq:Israeljunction1}-\eqref{eq:Israeljunction3} then fix the following three integration constants:
\begin{equation}
C_1^{f_2}=0, \qquad C_1^{A_2}= \frac{ e}{\pi } \sqrt{1-\eta}\,,\qquad
C_2^{A_2}= \frac{ e}{\pi }\sqrt{1-\eta} \left( 1-\text{Ci}(2 \pi )+ \frac{\pi }{ e} \,B_2^{A_2} \right).
\end{equation}

Consider now the scalar field equation at order $\mathcal O(\varepsilon^3)$, \ie still at $n=1$. The result for $\phi^{in}_3(R)$ is a long expression that is not very enlightening and hence we do not show it here. However, we give the Taylor expansions at the origin and at the box in order to show how we apply the boundary and junction conditions for the reader who wants to reproduce the results.

A Taylor series at the origin of $\phi^{in}_3(R)$ gives
\begin{align}
\phi^{in}_3(0)=\frac{1}{R}\left(\alpha_1+\frac{1}{12} \left(8-\frac{3 \left(e^2-2 \pi  \,\alpha_2\right)}{\pi ^2}\right)\right)+\mathcal O(R^0),
\end{align}
where $\alpha_1$ and $\alpha_2$ are the integration constants of the second order ODE for $\phi^{in}$. Regularity at the origin demands that we fix
\begin{align}
\alpha_2=\frac{e^2}{2 \pi }-\frac{2\pi}{3}  (3 \alpha_1+2).
\end{align}
We have two further conditions \eqref{def:epsilon}  at the box (vanishing of the field, and the definition of $\varepsilon$) that we use to fix the two remaining integration constants. 
Indeed recall that at the box one must have $\phi(R=1)=0$. However, one has $\phi^{in}_3(R=1)\neq 0$: 
\begin{align}\label{expphi3box0}
\phi^{in}_3(1)=\frac{e}{\pi}\,B_2^{A_2}+\frac{3}{2}-\frac{e^2}{\pi^2} \left[\text{Ci}(2 \pi)+1 \right]-\frac{8 \pi ^2-3 e^2}{6 \pi ^3}\left[ 2\,\text{Si}(2 \pi )- \text{Si}(4 \pi )\right]+\mathcal O(R-1)
\end{align}
where the function $\text{Si}(x)=\int_0^x \frac{\sin z}{z}\mathrm d z$ is the sine integral function. We can get $\phi^{in}_3(R=1)=0$ as desired by choosing appropriately  the integration constant $B_2^{A_2}$ that we had not fixed at order $\mathcal O(\varepsilon^2)$. This is a common feature of this perturbative expansion, an integration constant at order $\mathcal O(\varepsilon^{2n})$ for the gauge field is fixed at order $\mathcal O(\varepsilon^{2n+1})$. 
After this choice the leading term of the Taylor expansion of the scalar field $\varepsilon^3\phi^{in}_3$ about $R=1$ becomes:\footnote{The properties of the special functions present in  \eqref{expphi3box} guarantee that it is a real quantity.}
\begin{align}\label{expphi3box}
\varepsilon^3\phi^{in}_3(1)=& \,\varepsilon^3 (R-1)\Big[\text{Ci}(2 \pi ) \left(\frac{11}{6}-\frac{e^2}{2 \pi ^2}\right)+\frac{1}{12} \left(8 i \pi  (3 \,\alpha_1+1)+\frac{3 (2-i \pi ) e^2}{\pi ^2}-6\right) \nonumber\\
&+\text{Ci}(4 \pi ) \left(\frac{e^2}{2 \pi ^2}-\frac{4}{3}\right)
+\frac{8 \pi ^2-3 e^2}{12 \pi ^3} \left[2\text{Si}(2 \pi )-\text{Si}(4 \pi )\right] \Big] +\mathcal O(\varepsilon^3(R-1)^2).
\end{align}
However, recall from \eqref{def:epsilon} that we have defined unambiguously our expansion parameter $\varepsilon$ using the criterion that the expansion of the scalar field at the box is $\phi'(R=1)=\varepsilon$ at {\it all} orders in $\varepsilon$. This condition requires that \eqref{expphi3box} vanishes which fixes the remaining integration constant $\alpha_1$.   

At this point we have completely fixed the soliton up to order $\mathcal O(\varepsilon^3)$, \ie $n=1$, and confirmed that this is a regular asymptotically flat solution at this order.
A similar procedure can be used to extend the construction of the soliton to higher orders ($n>1$) in $\varepsilon$. In principle, we should be able to choose the integration constants at each order such that the solution is regular, \ie such that it obeys the boundary conditions at the origin \eqref{BCorigin} and asymptotically \eqref{BCinfinity} as well as the Israel junction conditions  \eqref{eq:Israeljunction1}-\eqref{eq:Israeljunction3}. To present the results \eqref{ThermoSoliton} below we have completed this exercise and determined the fields $f,A,\phi$ up to order $\mathcal{O}(\varepsilon^5)$ but we do not present the auxiliary computations/expressions because they are not further enlightening.  

Once we have obtained the fields $f, A, \phi$ up to order $\mathcal O(\varepsilon^5)$  we can compute the (gauge invariant) Brown-York quasilocal thermodynamic quantities at $\Sigma_{in}$ (defined in subsection \ref{subsec:quasilocal}) for the soliton  up to $\mathcal O(\varepsilon^6)$\footnote{Note that $f, A, \phi$ up to order $\mathcal O(\varepsilon^6)$ determines the chemical potential up to order $\mathcal O(\varepsilon^4)$.}.  As the soliton has no horizon, the entropy is zero and the temperature is undefined. On the other hand, the Brown-York quasilocal mass \eqref{BYmass}, quasilocal charge \eqref{BYcharge} and chemical potential $\mu\equiv A^{in}\big|_{R=1}$ are given by
\begin{align}\label{ThermoSoliton}
&{\cal M}/L=\varepsilon^2\, \frac{1}{2}+\varepsilon ^4 \, \frac{15 \pi ^2-6 e^2-16 \pi \left[2\, \text{Si}(2 \pi )-\text{Si}(4 \pi )\right]}{24 \pi ^2}+\mathcal O(\varepsilon^6),\nonumber\\
&{\cal Q}/L=\varepsilon^2\, \frac{e }{2 \pi }+\varepsilon ^4\,\frac{e}{8 \pi^4 }\Big[-\left(8 \pi ^2-e^2\right) \left[2\, \text{Si}(2 \pi )-\text{Si}(4 \pi )\right]+4 \pi  \left(e^2-2 \pi ^2\right)\Big]+\mathcal O(\varepsilon^6), \nonumber\\
&\mu=\frac{\pi }{e}+\varepsilon^2\,\frac{\left(8 \pi ^2-3 e^2\right) \left[2\, \text{Si}(2 \pi )-\text{Si}(4 \pi )\right]+3 \pi  \left(2 e^2-3 \pi ^2\right)}{6 \pi^2  e}+\mathcal O(\varepsilon^4).
\end{align}
As a non-trivial check of our computations, these thermodynamic quantities satisfy the quasilocal version of the first law of thermodynamics \eqref{1stlawsoliton}, $\mathrm d {\cal M}=\mu \,\mathrm d {\cal Q}$. At leading order, $\mu=\pi/e\equiv\tilde{\Omega}/e$, which corresponds to the lowest normal mode frequency $\tilde{\Omega}=\tilde{\omega} L=\pi$ that can fit inside the spherical box $\Sigma$ \cite{Dias:2018zjg}. 

The ADM mass and the charge measured by an asymptotic observer include the contribution associated to the energy-momentum tensor of the box $\Sigma$. That is to say, they depend on the constant $\eta$ that characterizes the matter content of the box; see discussion associated to \eqref{IsraelS0}. They are given by
\begin{align}\label{GlobalThermoSoliton}
&M/L=\frac{\eta}{2}+\varepsilon ^4 \,\frac{(1-\eta)e^2}{4\pi^2}+\mathcal O(\varepsilon^6),\\
&Q/L=\varepsilon ^2\, \frac{e\sqrt{1-\eta} }{2 \pi }+\varepsilon^4\,\frac{e\sqrt{1-\eta}}{8 \pi ^4 } \Big( 10 \pi ^3-4 \pi  e^2-\left(8 \pi ^2-e^2\right) \left[2 \,\text{Si}(2 \pi )-\text{Si}(4 \pi )\right]\Big)+\mathcal O(\varepsilon^6)\,.\nonumber
\end{align}

The Lanczos-Darmois-Israel surface energy-momentum tensor \eqref{eq:inducedT} at $\Sigma$ has non-vanishing components given by:
\begin{align}\label{IsraelSoliton}
&S^t_{\ t}=\frac{1}{8\pi}\Bigg[2\left(-1+\sqrt{1-\eta}\right)+\varepsilon ^2+\varepsilon^4\Bigg(\frac{5}{4}-\frac{e^2}{2\pi^2}-\frac{4 \left[2 \,\text{Si}(2 \pi )-\text{Si}(4 \pi )\right]}{3\pi }\Bigg)+\mathcal O\left(\varepsilon^6\right)\Bigg], \nonumber\\
&S^x_{\ x}=S^{\phi}_{\ \phi}=\frac{1}{8\pi}\Bigg[\left(\frac{1-\frac{1}{2}\eta}{\sqrt{1-\eta}}-1\right)-\frac{\varepsilon ^2}{2}+ \frac{\varepsilon ^4}{8}\left(1+\frac{2e^2}{\pi^2}\left(1-\sqrt{1-\eta}\right)\right)+\mathcal O\left(\varepsilon^6\right)\Bigg].
\end{align}
When $\varepsilon=0$, this reduces to \eqref{IsraelS0}. In this case,  the surface layer $\Sigma$ splits a Minkowski interior from a Schwarzschild exterior with the ADM mass  $M/L=\eta/2$ measuring the mass of the box.
In Section \ref{subsec:Boxstructure} we further discuss the physical interpretation of this energy-momentum tensor.

Finally note that under a gauge transformation \eqref{eq:U1gauge} we can move to a frame where the scalar field is complex, $\phi=|\phi|e^{-i\omega t}$, \ie where the scalar field oscillates in time with a frequency $\omega$. In this case the solitonic solution is often called a boson star. 

%%%%%%%%%%%%%%%%%%%%%%%%%%%
\section{Reissner-Nordstr\"om black hole confined in a box \label{sec:cagedRN}}
 
In Section  \ref{sec:HairyBH} we shall construct perturbatively an asymptotically flat hairy black hole solution whose scalar field is confined inside a cavity. In the absence of a horizon radius and scalar field, this cavity has no charge density but it has an energy density and pressure described by the Israel surface tensor \eqref{IsraelS0}, parametrized by the constant $\eta$. In the limit where the horizon radius vanishes, our hairy solution reduces to the hairy soliton found in Section \ref{sec:soliton}. On the other hand, when the scalar field is absent, the hairy black hole of  Section  \ref{sec:HairyBH}  reduces to a solution that describes a Reissner-Nordstr\"om black hole with horizon inside the Israel cavity. In this section we describe the gravitational and electric field of this solution. The interior solution and aspects of the thermodynamic properties of this solution (mainly in the grand-canonical ensemble) were already discussed in the seminal works \cite{Gibbons:1976pt,Hawking:1982dh,Braden:1990hw} and, more recently, in \cite{Andrade:2015gja,Basu:2016srp}. Here, for completeness, we also describe the fields in the region exterior to the box. For small energy and charge, the hairy black hole of Section  \ref{sec:HairyBH} can be seen as the solution that emerges from placing the small caged RN black hole of this section on top of the caged hairy soliton of Section \ref{sec:soliton}.

Consider first the region $R\leq 1$ inside the cavity located at $R=1$. The general solution to the equations of motion \eqref{eom} with the scalar field set to zero is given by:
\begin{equation}\label{RNin0}
f^{in}(R)=B_2^f-\frac{B_1^f}{R}+\frac{(B_1^A)^2}{2R^2}\,,
\qquad\qquad A^{in}(R)=B_2^A-\frac{B_1^A}{R}.
\end{equation}
Boundary conditions at the horizon \eqref{BChorizon} fix two integration constants,  $B_1^A=B_2^A R_+$ and $B_1^f=R_+\left(\frac{1}{2}(B_2^f)^2+B_2^f\right)$. At the box location, we use the scaling symmetry \eqref{2scalingsym} to require $f^{in}(1)=1$. We also introduce the chemical potential \eqref{eq:QLchp} as it appears in the quasilocal first law \eqref{1stlawBH}, \ie $\mu=A^{in}(1)-A^{in}(R_+)$. These two conditions fix the two remaining constants as $B_2^A=\frac{\mu}{1-R_+}$  and $B_2^f=\frac{2-R_+(2-\mu^2)}{2 (1-R_+)^2}$. Altogether, the fields of a RN BH whose horizon is confined inside a box located at $R=1$ are:
\begin{align}\label{RNin}
\begin{split}
&f^{in}(R)=\frac{2-R_+(2-\mu^2)}{2(1-R_+)^2}-\frac{R_+ \left[\mu^2 (1+R_+)+2 (1-R_+)\right]}{2 R (1-R_+)^2}+\frac{\mu^2 R_+^2}{2 R^2 (1-R_+)^2}, \\ 
&A^{in}(R)= \frac{\mu}{1-R_+}\left( 1-\frac{R_+}{R}\right), \qquad \qquad
g^{in}(R)=\frac{2-R_+(2-\mu^2)}{2(1-R_+)^2}\frac{1}{f^{in}(R)}\,.
\end{split}
\end{align}

Consider now the solution outside the box, $R\geq 1$. The most general solution reads
\begin{equation}\label{RNout0}
f^{out}(R)=C_2^f-\frac{C_1^f}{R}+\frac{(C_1^A)^2}{2R^2}\,,
\qquad\qquad A^{out}(R)=C_2^A-\frac{C_1^A}{R}.
\end{equation}
We use the scaling symmetry \eqref{2scalingsym} to set $f^{out}|_{R\to\infty}=1$ which fixes $C_2^f=1$. To fix the remaining integration constants we apply the Israel junction conditions \eqref{eq:Israeljunction1}-\eqref{eq:Israeljunction3} across the timelike hypersurface $\Sigma$. Moreover, we use the parametrizations \eqref{SigmaIn}-\eqref{SigmaOut} for the surface layer $\Sigma$.

We want to confine the horizon radius of the RN BH inside the `same cavity' that cages the hairy soliton of the previous section. By this statement we mean that the Israel surface tensor of the caged RN (and of our hairy solutions) reduces to the Israel surface tensor \eqref{IsraelS0} when $R_+\to 0$ (and/or $\varepsilon\to 0$).
This requires that we use exactly the same reparametrization factor $N=\left(1-\eta \right)^{-1/2}$ that was found in  \eqref{eq:solitonleadeingterms} and which describes the Israel surface tensor \eqref{IsraelS0} of our box.
This fixes  $C_1^f=\frac{\eta  \left[2 (1-R_+)^2-\mu ^2 R_+^2 \right]+\mu ^2  R_+^2}{2 (1- R_+)^2}$, $C_1^A= \mu   R_+\, \frac{\sqrt{1-\eta }}{1- R_+}$, and $C_2^A= \mu\,\frac{\sqrt{1-\eta } }{1- R_+}$.
The final solution outside the surface layer then reads 
%\begin{equation}\label{RNout}
%C_1^f=\frac{\eta  \left[2 (1-R_+)^2-\mu ^2 R_+^2 \right]+\mu ^2  R_+^2}{2 (1- R_+)^2}\,,\quad C_1^A= \mu   R_+\, \frac{\sqrt{1-\eta }}{1- R_+}\,,\quad  C_2^A= \mu\,\frac{\sqrt{1-\eta } }{1- R_+} \,.
%\end{equation}
\begin{align}\label{RNout}
\begin{split}
& f^{out}(R)=1-\frac{1}{R}\frac{\eta  \left[2 (1-R_+)^2-\mu ^2 R_+^2 \right]+\mu ^2  R_+^2}{2 (1- R_+)^2}+\frac{1}{R^2} \frac{(1-\eta ) \mu ^2 R_+^2}{2 (1-R_+)^2}\,,\\ 
&A^{out}(R)= \frac{\mu \,\sqrt{1-\eta }}{1-R_+}\left( 1-\frac{R_+}{R}\right), \qquad \qquad
g^{out}(R)=\frac{1}{f^{out}(R)}\,.
\end{split}
\end{align}
where $\eta$ is the parameter that describes the Israel surface tensor \eqref{IsraelS0} of our box (even before we place a scalar field or horizon inside it). In particular, if we set $R_+=0$, \eqref{RNin} and \eqref{RNout} reduce to \eqref{eq:solitonleadeingterms} with $\phi=0$. That is to say, the zero horizon radius limit of the caged RN BH solution \eqref{RNin} and \eqref{RNout} describes the same solution as the zero scalar field limit of the soliton \eqref{eq:solitonleadeingterms}. This common solution simply describes a surface layer cavity, placed in Minkowski space, with Israel surface tensor \eqref{IsraelS0}.\footnote{That is, when we set $R_+=0$ one gets $f^{in}(R)=1, \, A_t^{in}(R)=\mu$ and $f^{out}(R)=1-\eta/R, \, A_t^{out}(R)=\mu \sqrt{1-\eta }$. Note that the time reparametrization factor $N=\left(1-\eta\right)^{-1/2}$ guarantees that $f^{out}|_{R\to\infty}=1$. This setup (with $\mu=0$) is the textbook example of a Lanczos-Darmois-Israel  static thin shell separating Minkowski spacetime in the interior from the Schwarzschild geometry in the exterior with $\eta$ being proportional to the mass of the shell as measured by an asymptotic observer: see \eg section 3.10 of \cite{Poisson:2004} (with rotation $a=0$).} 
In these conditions, the cavity is the same and it will be consistent to place the small caged RN black hole of this section at the center of the caged hairy soliton of the previous section, as we do in Section \ref{sec:HairyBH}.

With the interior solution \eqref{RNin}, we can use the quasilocal formalism presented in Section \ref{subsec:quasilocal} to compute the quasilocal thermodynamic quantities of a caged RN black hole as a function of the horizon radius and the chemical potential:
\begin{align}\label{RNQLThermo}
\begin{split}
T_H L&=\frac{2-\mu^2}{4 \pi R_+ \sqrt{2} \sqrt{2-\left(2-\mu^2\right) R_+}},\qquad S/L^2 =\pi R_+^2. \\
{\cal M}/L&=1-\frac{(1-R_+)\sqrt{2}}{\sqrt{2-\left(2-\mu^2\right) R_+}},\qquad
{\cal Q}/L =\frac{\mu R_+}{\sqrt{2} \sqrt{2-\left(2-\mu^2\right) R_+}}.
\end{split}
\end{align}
Notice that extremal black holes have chemical potential $\mu=\sqrt 2$, hence RN BH inside a box exist in the region of parameters $ \mu\in [0,\sqrt2]$ and $R_+\in ]0,1[$. 
These quasilocal thermodynamic quantities obey the quasilocal first law \eqref{1stlawBH}. We  will further discuss the quasilocal thermodynamic properties of the caged RN BH in Section \ref{subsec:thermoanalysis}.

As an aside note, observe that a small horizon radius expansion of the RN quasilocal thermodynamics \eqref{RNQLThermo} yields, at leading order, 
\begin{align}\label{eq:RNboxleading}
&T_H L=\frac{1}{8\pi  R_+}(2-\mu^2)+\mathcal O(R_+), \qquad S_H/L^2=\pi R_+^2\,,\nonumber \\
&{\cal M}/L=\frac{R_+(2+\mu^2)}{4}+\mathcal O(R_+^2),\qquad {\cal Q}/L=\frac{\mu R_+}{2}+\mathcal O(R_+^2).
\end{align}
These leading order quantities coincide with the familiar ADM thermodynamic quantities of a RN black hole in the {\it absence} of a cavity. This is to be expected since taking the limit $R_+\equiv\frac{r_+}{L}\to 0$ implies $r_+\ll L$ which is the same as taking the limit $L\to\infty$ where the box is transported to the asymptotic region. In this limit we should indeed recover the ADM quantities of a RN black hole. Making contact with our companion paper \cite{Dias:2018zjg}, note that the non-interacting thermodynamic model of Section V of \cite{Dias:2018zjg} (which only captures the leading order thermodynamics of the system) makes use of this property.

For completeness, from \eqref{eq:thermogeneral} and the asymptotic behaviour of \eqref{RNout}, we can also read the  ADM mass and charge of the caged RN black hole system,  
\begin{equation}\label{thermoRN}
M/L=\frac{\eta}{2}+R_+^2\frac{\mu^2(1-\eta)}{4(1-R_+)^2}\,,\qquad Q/L=\frac{\mu  R_+ \sqrt{(1-\eta)}}{2 (1-R_+)}\,.
\end{equation}

%%%%%%%%%%%%%%%%%%%%%%%%%%
\section{Small hairy black holes confined in a box \label{sec:HairyBH}}

In Section \ref{sec:soliton} we found that our theory \eqref{action} has hairy soliton solutions in addition to the caged Reissner-Nordstr\"om black hole solutions of Section \ref{sec:cagedRN}.  In these conditions, as argued in section V of \cite{Dias:2018zjg}, the theory should admit a third solution $-$ a hairy black hole $-$ that in the small energy/charge limit can be thought of as placing a small caged Reissner-Nordstr\"om black hole on top of a hairy soliton.  Following this intuition, the leading order thermodynamics of these hairy black holes was computed (without solving the equations of motion) in \cite{Dias:2018zjg} using a simple thermodynamic model that assumes that the hairy black hole is a non-interacting mixture of two constituents: the soliton and the caged RN black hole. The two solutions can indeed be `merged' to yield a hairy black hole  as long as the two components are in thermodynamic equilibrium, \ie that they have the same chemical potential \cite{Dias:2018zjg} .

On the other hand, there is yet another argument in favour of the existence of hairy black hole solutions of  \eqref{action}. It is well established that RN black holes are unstable to the superradiant and near-horizon scalar condensation instabilities \cite{Herdeiro:2013pia,Hod:2013fvl,Degollado:2013bha,Hod:2014tqa,Li:2014gfg,Dolan:2015dha,Sanchis-Gual:2015lje,Basu:2016srp,Ponglertsakul:2016wae,Sanchis-Gual:2016tcm,Sanchis-Gual:2016ros,Hod:2016kpm,Fierro:2017fky,Dias:2018zjg}. It is then natural to expect that if we take such an unstable RN black hole as initial data, the system should time evolve towards a new solution that has scalar hair floating above the horizon, with the electric repulsion balancing the gravitational collapse. If so, this hairy black hole should have higher entropy (for a given mass and charge) than the initial system.  These hairy black holes should then be the endpoint of the superradiant instability of a RN black hole in a box. A time evolution study done recently confirmed that a Reissner-Nordstr\"om black hole perturbed with a scalar field indeed evolves towards a hairy black hole \cite{Herdeiro:2013pia,Degollado:2013bha,Sanchis-Gual:2015lje,Sanchis-Gual:2016tcm,Sanchis-Gual:2016ros}. 

In this section, we solve the equations of motion of \eqref{action} perturbatively to construct these static hairy black holes. We will confirm that, at least for small mass and charge, the intuition of the previous paragraphs is correct. Our hairy black holes should therefore be the endpoint of the superradiant instability of a RN black hole in a box. 

%%%%%%
\subsection{Setting up the perturbation problem}

Asymptotically flat hairy black holes of the Einstein-Maxwell theory confined inside a box must solve the equations of motion \eqref{eom}, subject to boundary conditions \eqref{BChorizon}, \eqref{BCinfinity} and Israel junction conditions \eqref{eq:Israeljunction1}-\eqref{eq:Israeljunction3}. We construct them perturbatively and find analytical expressions for the fields.
The hairy black holes of our theory are a two-parameter family of solutions that we can take to be the asymptotic scalar amplitude $\varepsilon$ defined as the derivative of the scalar field at the $R=1$ box $-$ see \eqref{def:epsilon} $-$ and the horizon radius $R_+$. Therefore, the perturbative construction requires a double expansion of the fields in powers of $\varepsilon$ and $R_+$. 
To be able to solve analytically the equations of motion \eqref{eom} we do a matched asymptotic expansion, similar to those done in a similar context in AdS backgrounds in
\cite{Basu:2010uz,Bhattacharyya:2010yg,Markeviciute:2016ivy,Dias:2011at,Dias:2011tj,Stotyn:2011ns}. However, here we also need to impose the junction conditions at the mirror located at $r=L$ ($R=1$). 
As for the soliton case, the box location divides the spacetime into the {\it inside} ($^{in}$; $r\leq L$) and {\it outside} ($^{out}$,  $r\geq L$) regions and we use the Israel junction conditions to match the fields of these two regions at the box hypersurface $r=L$  ($R=1$). But, this time, we further divide the inside region  ($^{in}$; $r\leq L$) into two sub-domains, namely the {\it near} region  $r_+\leq r\ll L$ ($^{in-near}$) and the {\it far} region where $r_+\ll r< L$  ($^{in-far}$). Considering small black holes that have $r_+/L \ll 1$, the near and far regions inside the box have an overlapping zone, $r_+ \ll r\ll L$. In this overlapping region, we can match the set of independent parameters that are generated by solving the perturbative equations of motion in each of the near and far regions.

We can start our perturbative construction. First note that the chemical potential of the hairy black hole should itself have a double expansion in powers of $\varepsilon$ and $R_+$, 
\begin{equation}\label{expansionChemPot}
\mu=\sum_{n\geq 0}\varepsilon^{2 n}\sum_{k\geq 0}R_+^{k}\mu_{2n,k}.
\end{equation}
Indeed, in Section \ref{sec:soliton}, we saw that the soliton is the back-reaction of a normal mode of the Minkowski box to higher orders. At leading order, the chemical potential of the  soliton is related by a gauge transformation to a normal mode frequency, but it is corrected at higher orders in the $\varepsilon$ expansion. We must allow similar corrections when the horizon radius expansion parameter is present. We shall construct the hairy black hole family whose zero-radius limit is the {\it ground state} soliton of Section \ref{sec:soliton} (so with lowest energy for a given charge).\footnote{A similar construction can be done for the excited hairy black holes.}   
In our analysis we take  $\varepsilon\ll 1$ and $R_+\ll 1$ and we assume that $\mathcal{O}(\varepsilon^2)\sim\mathcal{O}(R_+)$. It follows that terms with same $(n+k)$ contribute equally to the perturbative expansion, e.g.  $\mathcal{O}\left(\varepsilon^{0},R_+^2\right)\sim \mathcal{O}\left(\varepsilon^{2},R_+\right)\sim \mathcal{O}\left(\varepsilon^{4},R_+^0\right)$.

Outside the box, $R\geq 1$, the black hole can be considered to be a small perturbation in $\varepsilon$ and $R_+$ of Minkowski spacetime and the scalar field is required to vanish. In addition to \eqref{expansionChemPot}, the fields in this region (we use the superscript $^{out}$ to refer to this zone) have the double expansion
\begin{align}
\begin{split}\label{eq:expout}
&f^{out}(R)=\sum_{n\geq 0}\varepsilon^{2 n}\sum_{k\geq 0}R_+^{k}\,f_{2 n,k}^{out}(R),\qquad A^{out}(R)=\sum_{n\geq 0}\varepsilon^{2 n}\sum_{k\geq 0}R_+^{k}\,A_{2 n,k}^{out}(R),\\
&\phi^{out}(R)=0\,,
\end{split}
\end{align}
where we took into account that odd powers of $\varepsilon$ do not give corrections of the fields $f, \ A$.

In the far region inside the box, $R_+\ll R<1$, the fields of the hairy black hole still have a similar double expansion but this time the scalar field is also present (we use the superscript $^{in-far}$ to refer to this domain):
\begin{align}
\begin{split}\label{eq:expfar}
&f^{in-far}(R)=\sum_{n\geq 0}\varepsilon^{2 n}\sum_{k\geq 0}R_+^{k}\,f_{2 n,k}^{in-far}(R),\qquad A^{in-far}(R)=\sum_{n\geq 0}\varepsilon^{2 n}\sum_{k\geq 0}R_+^{k}\,A_{2 n,k}^{in-far}(R),\\
&\phi^{in-far}(R)=\sum_{n\geq 0}\varepsilon^{2 n+1}\sum_{k\geq 0}R_+^{k}\,\phi_{2 n+1,k}^{in-far}(R).
\end{split}
\end{align}
By construction, our hairy black hole solution has a smooth $R_+ \to 0$ limit where it reduces to the hairy soliton constructed in the previous section. To make this limit straightforward,  we use the scaling symmetry \eqref{2scalingsym} in each region to require that  
\begin{equation}\label{cfBH}
  f^{in-far}\big|_{R=1}=1\,,  \qquad f^{out}\big|_{R\to \infty}=1\,, 
\end{equation}
which is equivalent to \eqref{cf}.

The outside fields (denoted collectively by $Q^{out}$) must obey the boundary conditions  \eqref{BCinfinity} and these outside fields further have to be matched with the inside-far region fields $Q^{in-far}$ at the timelike hypersurface $\Sigma$ using the parametrizations \eqref{SigmaIn}-\eqref{SigmaOut} and the Israel junction conditions \eqref{eq:Israeljunction1}-\eqref{eq:Israeljunction3}. In particular, we are constructing our hairy black hole by placing the caged Reissner-Nordstr\"om black hole of Section \ref{sec:cagedRN}  on top of the soliton of Section \ref{sec:soliton}. Both of these constituents are confined inside the same cavity.  This means that we use exactly the same reparametrization factor $N=\left(1-\eta \right)^{-1/2}$ that was found in  \eqref{eq:solitonleadeingterms} and that describes the Israel surface tensor \eqref{IsraelS0} of our box.\footnote{Recall that  $\eta$ essentially describes the energy-momentum tensor of the box we choose to start with, \ie even before it receives corrections proportional to the scalar field amplitude $\varepsilon$ and horizon radius $R_+$. We cannot set it to zero or else the energy conditions are not obeyed, as discussed below \eqref{IsraelS0} and in \eqref{IsraelSoliton}. Our caged RN BHs and hairy solutions have a israel surface stress tensor that reduces to \eqref{IsraelS0} when $R_+\to 0$ and/or $\varepsilon\to 0$}

More concretely, we solve the equations at each order $\{n,k\}$ analytically (with the help of {\it Mathematica}) and find the {\it out} ({\it in-far}) fields up to a total of 4 (6) integration constants. The boundary conditions  \eqref{BCinfinity}  for the metric fix some constant(s). The Israel junction conditions  at the box \eqref{eq:Israeljunction1}-\eqref{eq:Israeljunction3}  fix extra integration constants. We are left with a few integration constants and a chemical potential coefficient of \eqref{expansionChemPot}, to be fixed by the matching with the inside-near region, as described next. 

We take the small radius $R$ limit of the inside-far fields, $Q^{in-far}(R)\big|_{R\ll 1}$ in order to prepare these fields to be matched with the inside-near region fields (collectively denoted by $Q^{in-near}$) to be discussed below. The fields $Q^{in-far}$ turn out to be divergent when $R\to 0$ as $\frac{R_+}{R}$. This gives an indication that the solutions do not hold for $R\sim R_+$, which is reasonable as we can no longer consider the black hole to be a small perturbation of the Minkowski spacetime at this scale. This justifies why the far-region analysis inside the box is valid only for $R\gg R_+$.  Also, it follows that in the far-region we can safely do a Taylor expansion in $R_+ \ll 1$ and $\varepsilon \ll 1$ since the large hierarchy of scales between the solution parameters $R_+,\varepsilon$ and $R$ guarantees that they do not compete.

Consider finally the inside-near region, $R_+\leq R\ll 1$. Here, Taylor expansions in $R_+ \ll 1$ and $\varepsilon \ll 1$ should be done with some caution since these small parameters can now be of similar order as $R$. This is closely connected with the fact that the inside-far region solution above breaks down when  $R/R_+\sim \mathcal{O}(1)$. This suggests that to proceed with the inside-near region analysis we should first introduce  new radial, $y$, and time, $\tau$, coordinates as  
\begin{align}\label{eq:neartransform}
 y=\frac{R}{R_+}, \qquad \tau= \frac{T}{R_+}.
\end{align} 
In this new frame, the inside-near region corresponds to $1 \leq y \ll R_+^{-1}$. If we further require that $R_+\ll 1$ (as is necessarily the case in our perturbative expansion) one concludes that the inside-near region corresponds to $R_+\ll 1\leq y$ (and $y\gg \varepsilon$). In particular, Taylor expansions in $R_+ \ll 1$ and $\varepsilon \ll 1$ can now be safely done since the radial coordinate $y$ and the black hole parameters $R_+,\varepsilon$ have a large hierarchy of scales\footnote{A key step for the success of the matching expansion procedure is that a factor of $R_+$ (one of the expansion parameters) is absorbed in the new coordinates \eqref{eq:neartransform}.}.
Further physical insight is gained if we rewrite the caged RN solution \eqref{RNin} in the new coordinate system \eqref{eq:neartransform}: 
\begin{align}\label{RNy}
&\mathrm d s^2=R_+^2\Big(-f(y)\mathrm d \tau^2+g(y)\mathrm d y^2+y^2\mathrm d \Omega_{2}^2\Big),\\
& \hspace{1cm}f(y)=\left(1-\frac{1}{y}\right)\frac{ 2-\left(2-\mu^2\right) R_+ -\frac{\mu^2}{y}}{2 (1-R_+)^2},\nonumber \\
&\hspace{1cm}A_{\tau}(y)=R_+\;A_{T}(y)=R_+\;\frac{\mu }{1-R_+}\left(1-\frac{1}{y}\right).\nonumber %\\
%& \hspace{1cm}g(y)=\frac{2-\left(2-\mu^2\right) R_+}{\left(1-\frac{1}{y}\right) \left( 2-\left(2-\mu^2\right) R_+-\frac{\mu^2}{y}\right)},\nonumber \\
%& \hspace{1cm}\phi(y)=0.\nonumber
\end{align}
The explicit factor of $R_+\ll 1$ in $A_{\tau}(y)$ shows that in the inside-near region the electric field is weak. Thus, at leading order, the gauge field is suppressed in the equations of motion and the system is to be seen as a small perturbation around the neutral solution. The same holds when we add a small scalar condensate to the system. The near fields of the hairy black hole thus have the double expansion (we use the superscript $^{in-near}$ to refer to this region):
\begin{align}
\begin{split}\label{eq:expnear}
&f^{in-near}(y)=\sum_{n\geq 0}\varepsilon^{2 n}\sum_{k\geq 0}R_+^{k}\,f_{2 n,k}^{in-near}(y),\qquad A^{in-near}(y)=\sum_{n\geq 0}\varepsilon^{2 n}\sum_{k\geq 0}R_+^{k}\,A_{2 n,k}^{in-near}(y),\\
&\phi^{in-near}(y)=\sum_{n\geq 0}\varepsilon^{2 n+1}\sum_{k\geq 0}R_+^{k}\,\phi_{2 n+1,k}^{in-near}(y).
\end{split}
\end{align}
In these conditions we can now solve analytically the equations of motion \eqref{eom}-\eqref{gequation} for $Q^{in-near}(y)$. One has to consider these equations with the change of variables \eqref{eq:neartransform} at each order in $\{n,k\}$. As  \eqref{eom}-\eqref{gequation} effectively describe a system of 3 ODEs for $f,\ A$ and $\phi$ at each order we have a total of 6 integration constants to be fixed. Three are determined by the boundary conditions at the horizon \eqref{BCorigin} (one for each field) and the other three constants are fixed by the matching with the inside-far region solution in the overlapping region $R_+\ll R\ll 1$. For this matching, we restore the coordinates $\{T,\ R\}$ and consider the large radius expansion of the inside-near fields, $Q^{in-near}(R)\big|_{R\gg 1}$ (with the expansion coefficients available at the given order). We find that these diverge as a power of $R$, which shows that the near region analysis  breaks down at $R\sim 1$. This explains why the near region analysis is valid only for $R\ll 1$. 
We can then match the two regions in powers of $\varepsilon$, $R_+$ and $R$ to fix the three integrations constants that were not yet determined.

In the end of the day, after imposing boundary, junction and matching conditions at each order, we are left only with a free parameter that characterizes the energy-momentum content of the box. 

The reader can find examples of the matching asymptotic expansion procedure described in the literature \cite{Basu:2010uz,Bhattacharyya:2010yg,Markeviciute:2016ivy,Dias:2011at,Dias:2011tj,Stotyn:2011ns,Dias:2015nua,Dias:2016pma}. To illustrate how to apply it in the case at hand, which has the novelty with respect to the previous references that we have to introduce Israel junction conditions, we show how the computation procedure is applied to the lower orders in the next subsection. 
The reader not interested in these technical details can move straightforwardly to subsection \ref{sec:thermoBH} where we present the final result for the thermodynamic quantities that describe the hairy black hole solutions of \eqref{eom}.

%%%%%%%%%%%%%%%%%%%%%%%%%%%%%%%%%%%%
\subsection{Construction of the solution using  matching expansion and junction conditions\label{App:ConstructionBH}}
%%%%%%%%%%%%%%%%%%%%%%%%%%%%%%%%%%%%

In this subsection we illustrate the general computational procedure of last subsection. Namely, we determine the expansion of the fields of the hairy black hole.
We insert the double expansion \eqref{eq:expout}  (outside region), \eqref{eq:expfar} (inside-far region) or \eqref{eq:expnear} (inside-near region) into the equations of motion \eqref{eom}-\eqref{gequation} and solve the resulting perturbed equations order by order.

%%%%%%%%%%%%%%%%%%%%%%%%%%%
\subsubsection{Matching asymptotic expansion at $\mathcal O\left(\varepsilon^0,R_+^k\right)$}
%%%%%%%%%%%%%%%%%%%%%%%%%%%

At lowest order, $\mathcal O\left(\varepsilon^0,R_+^k\right)$ in the scalar amplitude expansion, the scalar field is absent and the solution simply describes a Reissner-Nordstr\"om black hole placed inside a cavity with Israel surface stress tensor \eqref{IsraelS0}.

Consequently, the inside-far field coefficients $\{f^{in-far}_{0,k}(R),A^{in-far}_{0,k}(R)\}$ can be read directly from an $R_+\ll 1$ expansion of the  RN solution \eqref{RNin}, once we replace the chemical potential by its expansion \eqref{expansionChemPot}. For example, up to $\mathcal O\left(\varepsilon^0,R_+^2\right)$, this yields the solutions
\begin{eqnarray}\label{expBHinFar0}
&& f^{in-far}_{0,0}(R)=1\,, \qquad f^{in-far}_{0,1}(R)=-\frac{(1-R)\left(2+\mu_{0,0}^2\right) }{2 R}\,, \nonumber\\
&& \hspace{1.5cm} f^{in-far}_{0,2}(R)=\frac{(1-R)\left[\mu_{0,0}^2-2 R (1+\mu_{0,0}^2 +\mu_{0,0}\mu_{0,1})\right]}{2 R^2}\,; 
\nonumber\\
&&  A^{in-far}_{0,0}(R)=\mu_{0,0}\,, \qquad A^{in-far}_{0,1}(R)=\mu_{0,1}+\mu_{0,0}\left(1-\frac{1}{R}\right),\nonumber\\
&& \hspace{1.5cm} A^{in-far}_{0,2}(R)=\mu_{0,2}+\left(\mu_{0,0}+\mu_{0,1}\right)\left(1-\frac{1}{R}\right);
\end{eqnarray}
which satisfy the condition \eqref{cfBH} and the coefficients $\mu_{0,k}$ are to be determined by the matching conditions with the inside-near fields.

A similar Taylor expansion of \eqref{RNy} yields the inside-near field coefficients $\{f^{in-near}_{0,k}(y),A^{in-near}_{0,k}(y)\}$. Namely, up to $\mathcal O\left(\varepsilon^0,R_+^2\right)$, these are
\begin{eqnarray}\label{expBHinNear0}
&& \hspace{-0.4cm}f^{in-near}_{0,0}(y)=\left(1-\frac{1}{y}\right) \left(1-\frac{\mu_{0,0}^2}{2y}\right), 
\qquad f^{in-near}_{0,1}(y)=\left(1+\frac{\mu_{0,0}^2}{2}-\frac{\mu_{0,0}^2+\mu_{0,0} \mu_{0,1}}{y}\right)\left(1-\frac{1}{y}\right), \nonumber\\
&& \hspace{0.5cm} f^{in-near}_{0,2}(y)=\left(1+\mu_{0,0} (\mu_{0,0}+\mu_{0,1})-\frac{\mu_{0,1}^2+\mu_{0,0} (3 \mu_{0,0}+4 \mu_{0,1}+2 \mu_{0,2})}{2 y}\right)\left(1-\frac{1}{y}\right); \nonumber\\
&&  \hspace{-0.4cm}A^{in-far}_{0,0}(y)=\mu_{0,0} \left(1-\frac{1}{y}\right), \qquad A^{in-near}_{0,1}(R)=(\mu_{0,0}+\mu_{0,1})\left(1-\frac{1}{y}\right),\nonumber\\
&& \hspace{0.5cm}
 A^{in-near}_{0,2}(y)=\left(\mu_{0,0}+\mu_{0,1}+\mu_{0,2}\right)\left(1-\frac{1}{y}\right); 
\end{eqnarray}
and they obey the horizon boundary conditions \eqref{BChorizon}.
By construction (since we have just expanded the exact caged RN solution), the inside-far fields \eqref{expBHinFar0} and the inside-near fields \eqref{expBHinNear0} trivially satisfy the matching conditions in the overlapping region $R_+ \ll R\ll 1$.

Consider now the region outside the cavity, $R \geq 1$. Up to $\mathcal O\left(\varepsilon^0,R_+^2\right)$, the most general solution can be written as
\begin{eqnarray}\label{expBHout0}
&& f^{out}_{0,0}(R)=1-\frac{\tilde{\eta}}{R}\,, \qquad f^{out}_{0,1}(R)=-\frac{C_1^{f_{0,1}}}{R}\,,  
 \qquad f^{out}_{0,2}(R)=-\frac{C_1^{f_{0,2}}}{R}+\frac{\tilde{\mu}_{0,0}^2}{2 R^2} \,; \nonumber\\
&&  A^{out}_{0,0}(R)=\tilde{\mu}_{0,0}\,, \qquad A^{out}_{0,1}(R)=\tilde{\mu}_{0,1}+\tilde{\mu}_{0,0}\left(1-\frac{1}{R}\right), \nonumber\\
&& \hspace{1.5cm} A^{out}_{0,2}(R)=\tilde{\mu}_{0,2}+(\tilde{\mu}_{0,0}+\tilde{\mu}_{0,1})\left(1-\frac{1}{R}\right); 
\end{eqnarray}
where $\tilde{\eta}$, $\tilde{\mu}_{0,k}$,  and $C_1^{f_{0,k}}$ are integrations constants\footnote{Notice that we have already imposed the asymptotic boundary conditions \eqref{BCinfinity} (see also \eqref{cfBH}) to get \eqref{expBHout0}.}. They are determined applying the Israel junction conditions \eqref{eq:Israeljunction1}-\eqref{eq:Israeljunction3} across the timelike hypersurface $\Sigma$. We use the parametrizations \eqref{SigmaIn}-\eqref{SigmaOut} for the surface layer $\Sigma$ and, for reasons explained below \eqref{cfBH}, we use the same reparametrization factor $N=\left(1-\eta \right)^{-1/2}$ that was found in \eqref{eq:solitonleadeingterms}.
This fixes  
\begin{equation}
\tilde{\eta}=\eta\,, \qquad C_1^{f_{0,1}}=0,\qquad C_1^{f_{0,2}}=\frac{\tilde{\mu}_{0,1}^2}{2}\,,\qquad \tilde{\mu}_{0,k}= \mu_{0,k} \sqrt{1-\eta}\,,
\end{equation}
where $\eta$ is the parameter that describes the Israel surface tensor \eqref{IsraelS0} of our box (even before we place the scalar field and the horizon inside it). Again note that, when $R_+=0$, our system simply describes Minkowski spacetime with a constant electric field confined inside a surface layer with energy density described by $\eta$ (and no electric charge density). Consequently, in this case, the outside region is described by the Schwarzschild solution with ADM mass parameter $\eta$ that accounts for the mass of the surface layer.

%%%%%%%%%
\subsubsection{Matching asymptotic expansion at $\mathcal O\left(\varepsilon^1,R_+^k\right)$}\label{App:subsec:ConstructionBH1}
%%%%%%%%%

Moving to the next order in $\varepsilon$, the $\mathcal{O}(\varepsilon^1)$  correction switches on the scalar field $\phi$ without back-reacting it yet in the gravitoelectric background: it describes a small perturbation of the scalar field around the caged RN black hole. That is to say, the non-trivial equations of motion \eqref{eom}-\eqref{gequation}  reduce to the Klein-Gordon equation without a source\footnote{At higher orders, the equation of motion for $\phi$ still has the form of a Klein-Gordon equation but with an inhomogeneous term sourced by the lower order fields and their derivatives.}. 

At leading order, $\mathcal{O}(\varepsilon^1,R_+^0)$, the horizon radius does not contribute. It follows that the scalar field at this order is the same as the soliton scalar field inside the box \eqref{solitonorder0}. This means that,   
\begin{align}\label{solutionBH0}
\begin{split}
 \phi^{in-far}_{1,0}(R)=-\frac{\sin (\pi  R)}{\pi  R}\,,\qquad  \phi^{in-near}_{1,0}(y)=-1.
\end{split}
\end{align}
The analysis at this order also fixes $\mu_{0,0}=\frac{\pi}{e}$.

Next, we consider $\mathcal{O}(\varepsilon^1,R_+^1)$, \ie we  determine $\phi_{1,1}$. 

$\noindent \bullet$ {\tt Outside region, $R\geq1$:}

The scalar field vanishes outside the box in our setup at all orders, $\phi^{out}=0$. 

$\noindent \bullet$ {\tt Inside-Far region, $R_+\ll R\leq 1$:}

Nonlinear terms of the type $\phi_{1,0}f_{0,1}$, $\phi_{1,0}A_{0,1}$ and similar derivative combinations source a non-homogeneous contribution to the Klein-Gordon equation for $\phi^{in-far}_{1,1}(R)$. This equation can be solved analytically yielding   
\begin{eqnarray}\label{solutionInFar1}
&&\hspace{-0.4cm} \phi^{in-far}_{1,1}(R)=\frac{\beta_1 e^{-i \pi  R} - \beta_2 e^{i \pi  R}}{R}
-\frac{\cos (\pi  R)}{4 \pi  e^2 R}
\Big[4 \mu_{0,1} e^3 R-2\pi ^3 [\text{Ci}(2 \pi  R)-\ln(R)]-\left(\pi ^3-2 \pi  e^2\right)R\Big]\nonumber\\
&&\hspace{1.5cm}+\frac{\sin (\pi  R)}{8 \pi ^2 e^2 R^2}  \left[2\pi ^5 R- \pi ^3(2+R)-2 \pi  e^2 (2-R)+4 \mu_{0,1} e^3 R+4 \pi ^4 R \,\text{Si}(2 \pi  R)  \right]
\end{eqnarray}
where ${\rm Ci}(x)$ and ${\rm Si}(x)$ are (again) the cosine and sine integral functions and $\beta_{1,2}$ are two integration constants.

At the box location, the scalar field $\phi^{in-far}=\varepsilon \left(\phi_{0,0}^{in-far}+ \phi_{0,1}^{in-far} R_+ \right)$ has the Taylor expansion,
\begin{eqnarray}\label{solutionInFarBox}
&& \hspace{-0.5cm}\phi^{in-far}\big|_{R=1} \simeq \varepsilon \Big [ (R-1)+
R_+\left( -\beta_1+\beta_2 + \frac{e}{\pi}\,\mu_{0,1}-\frac{\pi ^2}{4 e^2}[1+2 \text{Ci}(2 \pi )]\right) \nonumber\\
&& +(R-1) R_+\left(\frac{2 e^2-2 \pi ^4+3 \pi ^2}{8 e^2}+\beta_1-\beta_2 +i \,\pi (\beta_1+\beta_2)-\frac{e\,\mu_{0,1}}{2 \pi }+\frac{\pi ^2}{2 e^2}[\text{Ci}(2 \pi )-\pi  \text{Si}(2 \pi )] \right) \nonumber\\
&& +\mathcal{O}\left((R-1)^2\right) \Big]. 
\end{eqnarray}
The junction conditions \eqref{def:epsilon} and associated choice for the definition of $\varepsilon$ requires that $\phi^{in-far}\big|_{R=1}= \varepsilon (R-1)+\mathcal{O}\left( (R-1)^2\right)$. Therefore the term proportional to $R_+$ in the first line of \eqref{solutionInFarBox} must vanish and so does the second line. These two conditions  fix $\beta_1$ and $\beta_2$.

$\noindent \bullet$ {\tt Inside-Near region, $R_+\ll R\ll 1$:}

Since at leading order the scalar field is a constant in this region, $\phi^{in-near}_{1,0}(y)=-1$, and the source term at order $\mathcal{O}(\varepsilon^1,R_+^1)$  depends only on the derivative of $\phi^{in-near}_{1,0}(y)$, the particular solution vanishes and  $\phi^{in-near}_{1,1}(y)$ is just the solution of the homogeneous Klein-Gordon equation. Regularity at the horizon requires that we set one of its two integration constants to zero, $\sigma_2=0$, to avoid a divergence $\ln(1-y)$. We are simply left with a constant solution
\begin{equation}
\phi^{in-near}_{1,1}(R)=\sigma_1\,.
\end{equation}

$\noindent \bullet$ {\tt Matching in overlapping region $R_+ \ll R \ll 1$ (inside the box):}

To do the matching we take the small $R$ limit of the inside-far region solution and the large $R$ limit of the inside-near solution. Note that at this order only terms up to $R^0 R_+^1$ and $R^1 R_+^0$ are appropriately accounted for (terms higher than this receive corrections from the next order). 
The small $R$ expansion of the inside-far region solution $\phi^{in-far}=\varepsilon \left(\phi_{0,0}^{in-far}+ \phi_{0,1}^{in-far} R_+ \right)$ is:
\begin{eqnarray}
\hspace{-1cm}\phi^{in-far}\big|_{R\ll 1}&\simeq&\varepsilon\Big[-1+ R_+\Big( \frac{\pi ^2 [1-2 \pi  \text{Si}(2 \pi )]}{4 e^2}+\frac{1}{2} \Big) \nonumber\\
&&+\frac{1}{2\pi e^2 R} \Big[ \pi ^3 [\gamma -1- \text{Ci}(2 \pi )+\ln (2 \pi )]+2\mu_{0,1} e^3 \Big]+\mathcal{O}\left( R^2,R_+R\right)\Big] 
\end{eqnarray}
while the inside-near region solution is simply $\phi^{in-near}_{1,1}(R)=\sigma_1$. Here, $\gamma\sim 0.577216$ is Euler's constant.
So, the far region solution breaks down for small radius (as expected from a previous discussion) and we can fix $\mu_{0,1}$ to eliminate  the divergent term $1/R$. Additionally, $\sigma_1$ is determined by matching the constant contributions. Altogether,
\begin{eqnarray}
\sigma_1 &=& \frac{1}{2}+\frac{\pi ^2 [1-2 \pi  \text{Si}(2 \pi )]}{4 e^2},\nonumber \\
\mu_{0,1}&=& \frac{\pi ^3 [1+\text{Ci}(2 \pi )-\gamma -\log (2 \pi )]}{2 e^3}.
\end{eqnarray}
We have fixed all the integration constants at order $\mathcal O(\varepsilon^1, R_+^1)$.

Should we wish, we could proceed with a similar procedure to orders $\mathcal{O}\left(\varepsilon,R_+^k\right)$ with $k\geq 2$ but this is not necessary for our physical purposes.

\subsubsection{Matching asymptotic expansion at $\mathcal O\left(\varepsilon^2,R_+^k\right)$\label{App:subsec:ConstructionBH2}}

At order $\mathcal O(\varepsilon^2)$, the $\mathcal O(\varepsilon)$ scalar field back-reacts on the metric and gauge field. The Klein-Gordon equation is trivially satisfied and we solve the remaining equations of motion for $f_{2,k}$ and $A_{2,k}$.

$\noindent \bullet$ {\tt Outside region, $R\geq1$:}

We solve the order $\mathcal O(\varepsilon^2, R_+^0)$ in the outer region for the fields $f^{out}_{2,0}$ and $A^{out}_{2,0}$ up to two integration constants for each field. Boundary conditions \eqref{BCinfinity} fix one integration constant and we are left with three undetermined integration constants $\{C_1^{f_{20}},C_1^{A_{20}},C_2^{A_{20}} \}$ that will be determined by the Israel junction conditions at the box:
\begin{equation}
f_{2,0}^{out}(R)=-\frac{C_1^{f_{20}}}{R}\,,\qquad A_{2,0}^{out}(R)=C_2^{A_{20}}-\frac{C_1^{A_{20}}}{R}\,.
\end{equation}

$\noindent \bullet$ {\tt Inside-Far region, $R_+\ll R\leq 1$:}

Also at order $\mathcal O(\varepsilon^2, R_+^0)$ we solve the far region equations for the fields $f^{in-far}_{2,0}$ and $A^{in-far}_{2,0}$. Each of the fields has a pair of integration constants, $\{B_1^{f_{20}},B_2^{f_{20}}\}$ and $\{B_1^{A_{20}},B_2^{A_{20}}\}$. The fields in this region are
\begin{align}
\begin{split}
& f_{2,0}^{in-far}(R)= B_1^{f_{20}}-\frac{B_2^{f_{20}}}{R}-2 \text{Ci}(2 \pi  R)+2 \ln R+\frac{\sin (2 \pi  R)}{\pi  R},\\
& A_{2,0}^{in-far}(R)=B_2^{A_{20}}-\frac{B_1^{A_{20}}}{R}+\frac{e}{2 \pi ^2 R}  \Big[ 2 \pi  R \Big(\ln R-\text{Ci}(2 \pi  R)\Big)+\sin (2 \pi  R)\Big].
\end{split}
\end{align}

These fields at the box take the values
\begin{align}
\begin{split}
f_{2,0}^{in-far}(1)=&B_1^{f_{20}}-B_2^{f_{20}}-2 \text{Ci}(2 \pi),\\
A_{2,0}^{in-far}(1)=&B_2^{A_{20}}-B_1^{A_{20}}-\frac{e \text{Ci}(2 \pi  )}{\pi }.
\end{split}
\end{align}
Conditions \eqref{cfBH} and  \eqref{eq:QLchp} allow to fix two integration constants, say $B_2^{A_{20}}=B_1^{A_{20}}+\frac{e}{\pi}\text{Ci}(2 \pi  )+\mu_{2,0}$ and $B_2^{f_{20}}=B_1^{f_{20}}-2 \text{Ci}(2 \pi)$. The other two are determined below by the matching conditions.

$\noindent \bullet$ {\tt Inside-Near region, $R_+\ll R\ll 1$:}

We solve the near region equations for the fields $f^{in-near}_{2,0}$ and $A^{in-near}_{2,0}$. Each of the fields has two integration constants, $\{K_1^{f_{20}},K_2^{f_{20}}\}$ and $\{K_1^{A_{20}},K_2^{A_{20}}\}$. We fix two of them, $K_2^{f_{20}}$ and $K_2^{A_{20}}$,  by imposing the boundary conditions at the horizon \eqref{BChorizon}. We obtain 
\begin{align}
\begin{split}
f_{2,0}^{in-near}(y)=&K_1^{f_{20}}-\frac{\pi K_1^{A_{20}}}{e y}\,\left(1-\frac{1}{y}\right),\\
A_{2,0}^{in-near}(y)=&K_1^{A_{20}}\,\left(1-\frac{1}{y}\right).
\end{split}
\end{align}
and $K_1^{f_{20}}, K_1^{A_{20}}$ are determined by the matching conditions that follow.

$\noindent \bullet$ {\tt Matching in overlapping region $R_+ \ll R \ll 1$ (inside the box):}

In order to match the inside-far and  inside-near regions, we take the large (small) $R$ expansion of the inside-near (inside-far) fields:
\begin{align}
\begin{split}
& f_{2,0}^{in-near}(y)\Big|_{R\gg 1}\simeq K_1^{f_{20}}+\mathcal O(R_+^1),\\
& A_{2,0}^{in-near}(y)\Big|_{R\gg 1}\simeq K_1^{A_{20}}+\mathcal O(R_+^1);\\
& f_{2,0}^{in-far}(R)\Big|_{R\ll 1}\simeq \frac{2\text{Ci}(2 \pi)-B_1^{f_{20}}}{R}+\big[B_1^{f_{20}}-2 \gamma +2-2 \ln (2 \pi )\big]+\mathcal O(R),\\
& A_{2,0}^{in-far}(R)\Big|_{R\ll 1}\simeq-\frac{ B_1^{A_{20}}}{R}+\frac{\pi  B_1^{A_{20}}-e \big[\gamma -1-\text{Ci}(2 \pi)+\ln (2\pi )\big]+\pi\mu_{2,0}}{\pi }+\mathcal O(R).
\end{split}
\end{align}
Note that we only keep terms that are not corrected by higher orders.
We fix 4 integration constants with the matching conditions,
\begin{eqnarray}
B_1^{f_{20}}=2\text{Ci}(2 \pi)&,\qquad &K_1^{f_{20}}= 2\text{Ci}(2 \pi)-2 \gamma +2-2 \ln (2 \pi ),  \qquad \\
B_1^{A_{20}}=0&, \qquad &K_1^{A_{20}}= \frac{\pi\mu_{2,0}-e [\gamma -1-\text{Ci}(2 \pi)+\ln (2\pi )]}{\pi }.
\end{eqnarray}

$\noindent \bullet$ {\tt Israel junction conditions at the box, $R=1$:}

The remaining constants are determined from the Israel junction conditions \eqref{eq:Israeljunction1}-\eqref{eq:Israeljunction3} across the timelike hypersurface $\Sigma$. Recall that in these conditions we use the time reparametrization introduced in \eqref{SigmaOut} and hence we take  $N=\left(1-\eta \right)^{-1/2}$ found in \eqref{eq:solitonleadeingterms}:
\begin{align}
&f^{in-far}_{2,0}(1)=0,\nonumber\\
&A^{in-far}_{2,0}(R)\Big|_{R\to 1}=\frac{\left(8 \pi ^2-3 e^2\right) \big[2 \text{Si}(2 \pi )-\text{Si}(4 \pi )\big]}{6 \pi ^2 e}+\frac{e}{\pi }-\frac{3 \pi }{2 e}+\frac{e }{\pi }(R-1)+\mathcal{O}\left((R-1)^2\right);\\
&N^2 f^{out}_{2,0}(1)= -\frac{C_1^{f_{20}}}{1-\eta}\,,\nonumber\\
&N A^{out}_{2,0}(R)\Big|_{R\to 1}=\frac{ C_2^{A_{20}}-C_1^{A_{20}}}{\sqrt{1-\eta}} + \frac{C_1^{A_{20}}}{\sqrt{1-\eta}}(R-1)+\mathcal{O}\left((R-1)^2\right).\nonumber
\end{align}
Continuity of the fields fixes the following integration constants:
\begin{align}
\begin{split}
&C_1^{f_{20}}=0,\qquad C_1^{A_{20}}= \sqrt{1-\eta}\,\frac{e}{\pi},\\
& C_2^{A_{20}}=\sqrt{1-\eta} \left(\frac{\left(8 \pi ^2-3 e^2\right) [2 \text{Si}(2 \pi )-\text{Si}(4 \pi )]}{6 \pi ^2 e}+\frac{2e}{\pi }-\frac{3 \pi }{2 e}\right).
\end{split}
\end{align}

At this stage we have fixed all the integration constants at order $\mathcal O(\varepsilon^2, R_+^0)$. We still need to fix the chemical potential $\mu_{2,0}$ which is fixed at order $\mathcal O(\varepsilon^3, R_+^0)$ when the scalar field is found. In general in this perturbative scheme, the chemical potential $\mu_{n,k}$ is fixed at order $\mathcal O(\varepsilon^{n+1}, R_+^k)$.

%%%%%%%%%
\subsubsection{Matching asymptotic expansion at higher orders}

To arrive to the relevant physical results we also need to compute the correction to the scalar field at order $\mathcal{O}\left(\varepsilon^3\right)$ and to the gravitational and gauge field at order $\mathcal{O}\left(\varepsilon^2 R_+\right)$ and $\mathcal{O}\left(\varepsilon^4\right)$.
The computation of the former correction parallels that of Section \ref{App:subsec:ConstructionBH1} while the determination of the latter parallels the discussion of Section \ref{App:subsec:ConstructionBH2}. We therefore omit the details that are not further enlightening.

%%%%%%%%%%%%%%%%%%%%%%%%%%%
\subsection{Thermodynamic quantities\label{sec:thermoBH}}
%%%%%%%%%%%%%%%%%%%%%%%%%%%
With the perturbation expansion presented in the previous subsection we have found the scalar, gauge and gravitational fields perturbatively. We can now insert these fields into the expressions for the quasilocal thermodynamics of subsection \ref{subsec:quasilocal} to compute the thermodynamic quantities of the hairy BH. In the perturbation expansion we have assumed that $\mathcal{O}(\varepsilon^2)\sim\mathcal{O}(R_+)$. This assumption implies that terms with the same $(n+k)$ contribute equally to the perturbative expansion, \eg  $\mathcal{O}\left(\varepsilon^{0},R_+^2\right)\sim \mathcal{O}\left(\varepsilon^{2},R_+\right)\sim \mathcal{O}\left(\varepsilon^{4},R_+^0\right)$. We did this consistent perturbative expansion up to the order necessary to get the thermodynamic quantities that verify the first law of thermodynamics up to order $(n+k)=2$. This is enough to get our main result best summarized in later Fig. \ref{fig:MCphaseDiag}.

Using \eqref{eq:thermogeneral}, the dimensionless quasilocal energy $\cal M$, quasilocal charge $\cal Q$, chemical potential $\mu$, temperature $T$ and entropy of the small hairy black holes are given by:
\begin{align}\label{ThermoHairyBH}
{\cal M}/L=&\Bigg[\frac{R_+}{4} \left(\frac{\pi ^2}{e^2}+2\right) +\frac{R_+^2}{32 e^4} \Bigg(\pi ^4 \Big(8 [\text{Ci}(2 \pi )- \gamma-\ln (2 \pi )] +5\Big)+4 \left(e^2+\pi ^2\right) e^2\Bigg)+\mathcal{O}(R_+^3)\Bigg]\nonumber\\
&+\varepsilon ^2 \Bigg[\frac{1}{2}+\frac{R_+}{12 \pi  e^2} \Bigg( 9 \pi ^3 \Big[\gamma-\text{Ci}(2 \pi ) -2+\ln (2 \pi )\Big]+\left(8 \pi ^2-3 e^2\right) \Big[ 2 \text{Si}(2 \pi )-\text{Si}(4 \pi )\Big] \Bigg)\nonumber\\
&+\mathcal{O}(R_+^2)\Bigg]+\varepsilon ^4 \Bigg[\frac{15 \pi ^2-6 e^2+16 \pi  \big[\text{Si}(4 \pi )-2 \text{Si}(2 \pi )\big]}{24 \pi ^2}+\mathcal{O}(R_+)\Bigg]+\mathcal{O}(\varepsilon^6)\nonumber\\
{\cal Q}/L=&\Bigg[\frac{\pi  R_+}{2 e}+\frac{R_+^2}{8 e^3} \Bigg(\pi ^3 \Big(2 [\text{Ci}(2 \pi )-\gamma - \ln (2 \pi )]+1\Big)+2 \pi  e^2\Bigg)+\mathcal{O}(R_+^3)\Bigg]+\varepsilon ^2 \Bigg[\frac{e}{2 \pi }\nonumber\\
&\hspace{-0.4cm}+\frac{R_+}{12 \pi ^2 e} \Bigg(12 \pi ^3 \Big(\gamma-\text{Ci}(2 \pi ) +\ln (2 \pi )-\frac{7}{4}\Big)+\left(8 \pi ^2-3 e^2\right) \big[2 \text{Si}(2 \pi )-\text{Si}(4 \pi )\big]\Bigg)+\mathcal{O}(R_+^2)\Bigg]\nonumber\\
&\hspace{-0.4cm}-\Bigg[\varepsilon ^4\frac{e \left(\left(8 \pi ^2-e^2\right) (2 \text{Si}(2 \pi )-\text{Si}(4 \pi ))+4 \pi  e^2-8 \pi ^3\right)}{8 \pi ^4}+\mathcal{O}(R_+)\Bigg]+\mathcal{O}(\varepsilon^6)\nonumber\\
& \hspace{-1.1cm}\mu=\Bigg[\frac{\pi }{e}-R_+\,\frac{\pi ^3}{2 e^3} \Big(\gamma-\text{Ci}(2 \pi ) -1+\ln (2 \pi ) \Big)+\mathcal O(R_+^2)\Bigg]\nonumber\\
&\hspace{-0.3cm}+\varepsilon^2 \Bigg[\frac{8 \pi ^2-3 e^2}{6 \pi ^2 e} \Big(2 \text{Si}(2 \pi )-\text{Si}(4 \pi )\Big)+\frac{e}{\pi}-\frac{3 \pi}{2 e}+\mathcal O(R_+)\Bigg]+\mathcal O(\varepsilon^4),
\end{align}
\begin{align}
T L=&\frac{1}{4 \pi R_+}\Bigg\{\Bigg[1-\frac{\pi ^2}{2 e^2}+\frac{R_+}{8 e^4}\Bigg(4\pi ^4 \Big(\gamma-\text{Ci}(2 \pi ) +\ln (2 \pi )-\frac{3}{4}\Big)+4 \left(e^2-\pi ^2\right) e^2\Bigg)+\mathcal{O}(R_+^2)\Bigg]\nonumber\\
&+\varepsilon ^2\Bigg[ \frac{3 e^2-8 \pi ^2}{6 \pi  e^2} \Big(2 \text{Si}(2 \pi )-\text{Si}(4 \pi )\Big)-\frac{\pi ^2}{2  e^2} \Big(\gamma-\text{Ci}(2 \pi ) -4+\ln (2 \pi )\Big)-1+\mathcal{O}(R_+)\Bigg]\nonumber\\
&+\mathcal O(\varepsilon^4)\Bigg\},\nonumber\\
S/L^2=&\pi R_+^2.\nonumber
\end{align}
Recall that  $\text{Ci}(x)=-\int_x^{\infty}\frac{\cos z}{z}\mathrm d z$ and $\text{Si}(x)=\int_0^x \frac{\sin z}{z}\mathrm d z$  are the cosine and sine integral functions, respectively, and $\gamma\sim 0.577216$ is Euler's constant.

As an important check of the computations, notice that these thermodynamic quantities satisfy the quasilocal form of the first law of thermodynamics \eqref{1stlawBH}. Moreover, when $R_+=0$, \eqref{ThermoHairyBH} reduces to the soliton thermodynamics \eqref{ThermoSoliton} and, when $\varepsilon=0$, \eqref{ThermoHairyBH} yields the caged RN black hole thermodynamics \eqref{RNQLThermo} (once we insert $\mu$ in \eqref{ThermoHairyBH} into \eqref{RNQLThermo} and take a series expansion up to $\mathcal{O}(R_+^2)$).

From these thermodynamic quantities  \eqref{ThermoHairyBH}, we can confirm that the  simple non-interacting thermodynamic model presented in Section V of \cite{Dias:2018zjg} $-$ which does {\it not} use the equations of motion $-$ captures the correct {\it leading} order thermodynamics, \ie the leading terms of \eqref{ThermoHairyBH}.
First note that, at leading order, it follows from \eqref{ThermoHairyBH} that $\mu=\frac{\pi}{e}$.
Assuming, as justified above, that $\mathcal{O}(\varepsilon^2)\sim\mathcal{O}(R_+)$, the leading order contributions of the expansion \eqref{ThermoHairyBH} allows to express analytically $R_+$ and $\varepsilon$ in terms of ${\cal M}$ and ${\cal Q}$,
\begin{align}
&R_+= \frac{4 e (e {\cal M}-\pi  {\cal Q})}{2 e^2-\pi ^2}+\mathcal O\left({\cal M}^2,{\cal Q}^2,{\cal M} {\cal Q}\right), \quad \nonumber\\
&\varepsilon^2 =\frac{2 \pi  \left(2 e^2 {\cal Q}-2 \pi  e {\cal M}+\pi ^2 {\cal Q}\right)}{2 e^3-\pi ^2 e}+\mathcal O\left({\cal M}^2,{\cal Q}^2,{\cal M} {\cal Q}\right),
\end{align}
which we insert in the expressions for the other thermodynamic quantities to find that at leading order in ${\cal M}$ and ${\cal Q}$ one has:
\begin{align}\label{eq:ThermoLeadingOrder}
\begin{split}
& \mu=\frac \pi e+\mathcal O\left({\cal M}, {\cal Q}\right), \\
&S/L^2=\frac{16 \pi  e^2 (e {\cal M}-\pi  {\cal Q})^2}{\left(\pi ^2-2 e^2\right)^2}+\mathcal O\left({\cal M}^2,{\cal Q}^2,{\cal M} {\cal Q}\right),\\
&T L=\frac{\left(\pi ^2-2 e^2\right)^2}{32 \pi  e^3 (e {\cal M}-\pi  {\cal Q})}+\mathcal O\left({\cal M}^2,{\cal Q}^2,{\cal M} {\cal Q}\right).
\end{split}
\end{align}
These quantities \eqref{eq:ThermoLeadingOrder} match the result of the non-interacting model presented in equation (5.4) of Section V of \cite{Dias:2018zjg}. Thus the non-interacting thermodynamic model does indeed capture the leading order properties of this hairy black hole system at very low cost. This explicit confirmation adds to those done in similar hairy black hole systems in \cite{Basu:2010uz,Bhattacharyya:2010yg,Markeviciute:2016ivy,Dias:2011at,Dias:2011tj,Cardoso:2013pza,Dias:2015rxy,Dias:2016pma} and realize in simple terms the idea that we can place a small black hole, without hair, of the theory at the core of its hairy soliton to get the hairy black hole of the theory. 

For completeness, note that an asymptotic observer measures the ADM mass and ADM charge of the hairy black hole to be:\footnote{The constant $\mathcal{C}_1$ in the expression for $Q$ is given by: 
\footnotesize{
 \begin{align*}
 \mathcal{C}_1=&\frac{1}{8} \pi  \Bigg[2 i \pi  \left(G_{2,3}^{3,1}\left(-2 i \pi \left|
\begin{array}{c}
 -1,1 \\
 -1,0,0 \\
\end{array}
\right.\right)-G_{2,3}^{3,1}\left(2 i \pi \left|
\begin{array}{c}
 -1,1 \\
 -1,0,0 \\
\end{array}
\right.\right)\right)-G_{2,3}^{3,1}\left(-2 i \pi \left|
\begin{array}{c}
 0,2 \\
 0,0,1 \\
\end{array}
\right.\right)\\
&-G_{2,3}^{3,1}\left(2 i \pi \left|
\begin{array}{c}
 0,2 \\
 0,0,1 \\
\end{array}
\right.\right)\Bigg]
\simeq   -0.0177191,\quad \hbox{where}  \:\: G^{m\, n}_{p\, q} \left(x\left|
\begin{array}{c}
 a_1,\dots,a_p \\
 b_1,\dots, b_q \\
\end{array}
\right.\right) \:\: \hbox{is the MeijerG function.}
 \end{align*} }
These ADM quantities should obey a first law of thermodynamics that includes a term that accounts for the thermodynamic contribution of the Israel surface layer.  
}  
\begin{align}\label{GlobalThermoBH}
M/L=&\Bigg[\frac{\eta}{2}+R_+^2\frac{\pi^2(1-\eta)}{4e^2}+\mathcal{O}(R_+^3)\Bigg]+\varepsilon ^2\left(R_+ \frac{1-\eta}{2}+\mathcal{O}(R_+^2)\right)\nonumber\\
&+\varepsilon ^4\left( \frac{e^2(1-\eta)}{4\pi^2}+\mathcal{O}(R_+)\right)+\mathcal O(\varepsilon^6),\\
Q/L=&\Bigg[ R_+\sqrt{1-\eta}\frac{\pi}{2e}+R_+^2\frac{\pi  \sqrt{1-\eta} }{4 e^3}\Big(2 e^2+\pi ^2 \big[1-\gamma+\text{Ci}(2 \pi )-\ln (2 \pi )\big]\Big)+\mathcal{O}(R_+^3)\Bigg]\nonumber\\
&+\varepsilon^2 \Bigg[\frac{e\sqrt{1-\eta}}{2 \pi }+R_+\Bigg(\frac{\sqrt{1-\eta}}{24 \pi ^2 e} \Big(6 \pi  e^2+3 \pi ^3 \big[8 \gamma-10 \text{Ci}(2 \pi ) -11+8 \ln (2 \pi )\big] \nonumber\\
&\hspace{4.6cm}+(16 \pi ^2-6 e^2) \big[2 \text{Si}(2 \pi )-\text{Si}(4 \pi )\big]\Big)+ \frac{\sqrt{1-\eta}}{e}\mathcal{C}_1\Bigg)+\mathcal{O}(R_+^2)\Bigg]\nonumber\\
&+\varepsilon^4\Bigg[ \frac{e\sqrt{1-\eta} }{8 \pi ^4} \Big(\left(e^2-8 \pi ^2\right) \big[2 \text{Si}(2 \pi )-\text{Si}(4 \pi )\big]-4 \pi  e^2+10 \pi ^3\Big)+\mathcal{O}(R_+)\Bigg]+\mathcal O(\varepsilon^6).\nonumber
\end{align}

Finally, the  Lanczos-Darmois-Israel surface energy-momentum tensor \eqref{eq:inducedT} at the surface layer $\Sigma$ has non-vanishing components given by:
\begin{align}\label{IsraelHairyBH}
&S^t_{\ t}=\frac{1}{8\pi}\Bigg\{\Bigg[-2\left(1-\sqrt{1-\eta}\right)+R_+\left(1+\frac{\pi^2}{2e^2}\right)+R_+^2\Bigg(\frac{e^2+\pi ^2}{4 e^2}+\frac{\pi ^4}{16 e^4}\Big( 5-8 \big[\gamma-\text{Ci}(2 \pi ) \nonumber\\
&\hphantom{S^t_{\ t}=}+\ln (2 \pi )\big]\Big)\Bigg)+\mathcal O(R_+^3)\Bigg]+\varepsilon ^2\Bigg[1+R_+\Bigg(\frac{8 \pi ^2-3 e^2}{6 \pi  e^2} \big[2 \text{Si}(2 \pi )-\text{Si}(4 \pi )\big]+\frac{3 \pi ^2}{2 e^2}\Big(\gamma-\text{Ci}(2 \pi ) \nonumber\\
&\hphantom{S^t_{\ t}=}+\ln (2 \pi )-2\Big)\Bigg)+\mathcal O(R_+^2)\Bigg]+\varepsilon^4\Bigg[\frac{5}{4}-\frac{e^2}{2\pi^2}-\frac{4 \big[2 \text{Si}(2 \pi )-\text{Si}(4 \pi )\big]}{3\pi }\Bigg]+\mathcal O\left(\varepsilon^6\right)\Bigg\},\\
&S^x_{\ x}=S^{\phi}_{\ \phi}=\frac{1}{8\pi}\Bigg\{ \Bigg[\frac{2\left(1-\sqrt{1-\eta}\right)-\eta}{2 \sqrt{1-\eta}}-R_+^2\Bigg(\frac{\left(\pi ^2-2 e^2\right)^2}{32 e^4}+\frac{\pi^2}{8 e^2}\sqrt{1-\eta}\Bigg)+\mathcal O(R_+^3)\Bigg]\nonumber\\
&\hphantom{S^t_{\ t}=}+\frac{\varepsilon ^2}{2}\Bigg[-1+R_+\Big(1-\sqrt{1-\eta}\Big)+\mathcal O(R_+^2)\Bigg]+ \frac{\varepsilon ^4}{8}\Bigg[1+\frac{2 e^2}{\pi^2}\left(1-\sqrt{1-\eta}\right)\Bigg] +\mathcal O\left(\varepsilon^6\right)\Bigg\}.\nonumber
\end{align}
When $R_+=0$ this reduces to \eqref{IsraelSoliton}, as it should.

%%%%%%%%%%%%%%%%%
\section{Discussion of physical properties}\label{sec:Physprop}

%%%%%%%%%%%%%%%%%%%%%%%%%%%%%%%%%
%%%%%%%%%%%%%%%%%%%%%%%%%%%%%%%%%
\subsection{Israel surface stress tensor and energy conditions}\label{subsec:Boxstructure}
 
Scalar fields confined inside a box were already studied in the literature: 1) at the linear level \cite{Herdeiro:2013pia,Hod:2013fvl,Degollado:2013bha,Hod:2014tqa,Li:2014gfg,Hod:2016kpm,Fierro:2017fky,Li:2014xxa,Li:2014fna,Li:2015mqa,Li:2015bfa}, 2) as a nonlinear elliptic problem (although without having flat asymptotics \cite{Dolan:2015dha,Ponglertsakul:2016wae,Ponglertsakul:2016anb} or without discussing the exterior solution \cite{Basu:2016srp}), and 3) as an initial-value problem \cite{Sanchis-Gual:2015lje,Sanchis-Gual:2016tcm,Sanchis-Gual:2016ros}. However, to the best of our knowledge, the properties of the ``internal structure" of the cavity or surface layer that is necessary to confine the scalar field were never analysed. That is to say, it was never discussed whether the energy-momentum content of the mirror necessary to confine the scalar field is physically acceptable and, if so, what are its properties. Now that we have the regular and asymptotically flat hairy solitons and hairy black holes of the theory we can analyse the Lanczos-Darmois-Israel surface stress tensor  \eqref{eq:inducedT} and address these questions. At the frontline, we have to ask whether it does obey or not the energy conditions.

As pointed out before, in our construction we have imposed the Israel junction conditions on the gravitoelectric fields on the surface layer $\Sigma$. We decided to impose that the component of the electric field orthogonal to $\Sigma$ is continuous across $\sigma$. This amounts to have no surface electric charge density, which motivated our choice. With these conditions on $\Sigma$ the system is still left with a free parameter. Essentially we have a 1-parameter family of solutions, that is labeled by the parameter $\eta$, that differ on the energy-momentum content of the surface layer $\Sigma$ or, equivalently, on the total mass of the solution. That is to say, $\eta$ determines the mass of the shell and thus of the hairy solution (as can be checked inspecting the mass formulas in \eqref{GlobalThermoSoliton} and \eqref{GlobalThermoBH}). Different choices of $\eta$ dictate a distinct extrinsic curvature jump at $\Sigma$, \ie a different Lanczos-Darmois-Israel surface stress tensor. 

In \eqref{IsraelSoliton} and \eqref{IsraelHairyBH} we have computed the three non-vanishing components of the tensor $S^a_{\ b}$ for the soliton and hairy black hole, respectively. As a consequence of the conservation law for the Brown-York stress tensor (see subsection \ref{subsec:quasilocal}), the  Lanczos-Darmois-Israel surface tensor is also conserved, ${\cal D}_a S^{ab}=0$, as we can explicitly check. Further note that the zero horizon radius limit of the hairy black hole is the soliton and thus  that \eqref{IsraelHairyBH}  reduces to \eqref{IsraelSoliton} when we set $R_+\to 0$.

The  Lanczos-Darmois-Israel surface tensor  can be written in the perfect fluid form, ${\cal S}_{(a)(b)}= \mathcal{E} u_{(a)}u_{(b)}+ \mathcal{P}(h_{(a)(b)}+u_{(a)}u_{(b)})$, with $u=f^{-1/2}\partial_t$ and local energy density $ \mathcal{E}=[\rho]$ and pressure $ \mathcal{P}=[p]$ given by\footnote{These densities can also be computed following a different, but equivalent, route (see \eg \cite{Wald:106274}). Introduce the orthogonal tetrad on $\Sigma$, $e_{(0)}=f^{-1/2}\partial_t,\: e_{(k)}=r^{-1} \hat{e}_{(k)},\:\: (k=1,2),$
such that $h_{uv}e_{(a)}^{\phantom{(a)}u} e_{(b)}^{\phantom{(a)}v}=\eta_{(a)(b)}$ with $\eta_{(a)(b)}={\rm diag}\{-1,1,1 \}$ being the 3-dimensional Minkowski metric and $\hat{e}_{(k)\,i} \hat{e}^{(k)}_{\phantom{(k)} j}=\sigma_{ij}$. The components of the  Lanczos-Darmois-Israel energy-momentum tensor \eqref{eq:inducedT} (at $\Sigma$) in the tetrad frame are then ${\cal S}_{(a)(b)}=e_{(a)}^{\phantom{(a)}u} e_{(b)}^{\phantom{(a)}v} {\cal S}_{uv}$ and one has ${\cal S}_{(a)(b)}e^{(b)}=\lambda\, e_{(a)}$ which is an eigenvalue equation. That is, $\lambda=\{-\mathcal{E},\mathcal{P}_{i} \}$ are the eigenvalues (with $\mathcal{E}$ the energy density and $\mathcal{P}_{i}$ the two principal pressures) and $e_{(a)}$ are the associated unit eigenvectors (principal directions). In particular, $e_{(0)}$ is the 3-velocity of a box observer in her local rest frame.
}
\begin{equation}\label{densitiesbox}
\mathcal{E}=-S^t_{\ t}\,,\qquad \mathcal{P}=S^x_{\ x}=S^{\phi}_{\ \phi},
 \end{equation}
with $S^t_{\ t},S^x_{\ x},S^{\phi}_{\ \phi}$ defined in \eqref{IsraelSoliton} and \eqref{IsraelHairyBH} for the soliton and hairy black hole respectively.

Physical surface layers must have a stress tensor (\ie an energy density and pressure) that obeys the energy conditions. Different versions of these energy conditions read $(i=1,2)$ \cite{Wald:106274}:
\begin{eqnarray}\label{energyconditions}
\hbox{Weak energy condition:} && {\cal E}\geq 0 \quad \land \quad {\cal E}+{\cal P}_{i}\geq 0  \,;\\
\hbox{Strong energy condition:} & &{\cal E}+{\cal P}_{i}\geq 0 \quad \land \quad {\cal E}+\sum_{i=1}^2 {\cal P}_{i}\geq 0\,;\\
\hbox{Null energy condition:} & &{\cal E}+{\cal P}_{i}\geq 0\,;\\
\hbox{Dominant energy condition:}& & {\cal E}+|{\cal P}_{i}|\geq 0\,,
\end{eqnarray}

Set $R_+=0$ and $\varepsilon=0$ in \eqref{IsraelHairyBH} and \eqref{densitiesbox}.
A physical surface layer with $0<\eta\leq 1$ has an energy density and pressure and associated jump in the extrinsic curvature. Solutions with these parameters  split the spacetime into an interior region that is Minkowski spacetime and an exterior region which is described by the Schwarzschild geometry with ADM mass proportional to $\eta$, $M=\eta L/2$. If we choose $\eta=0$, the surface layer is  absent.   

Now, consider the hairy solitons  whose surface layer $\Sigma$ has stress tensor \eqref{IsraelSoliton}, \ie let $\varepsilon\neq 0$ but $R_+=0$ in \eqref{IsraelHairyBH} and \eqref{densitiesbox}. An inspection of these quantities quickly concludes that for $\eta=0$, none of the energy conditions \eqref{energyconditions} is obeyed. As discussed already below \eqref{IsraelS0}, we must have $\eta\neq 0$ to guarantee that they (or a relevant subset of them) are obeyed. Actually, we can find the minimum shell parameter $\eta$, as a power expansion in $\varepsilon$, above which (with $\eta<1$) the  energy conditions (or a select subset of them) are obeyed.
Consider now the hairy black hole with $\varepsilon\neq 0$ and $R_+\neq 0$ in \eqref{IsraelHairyBH} and \eqref{densitiesbox}. There is still a minimum shell parameter $\eta$, which we can express as a double power expansion in $\varepsilon$ and $R_+$, above which (with $\eta<1$) the energy conditions (or a select subset of them) are obeyed. 
To conclude this discussion on the energy conditions, as an important byproduct of our study, we have found the energy-momentum content  that a mirror must have to confine a hairy soliton or hairy black hole inside of it.  

%%%%%%%%%% 
 \subsection{Phase diagram of asymptotically flat solutions in a box}\label{subsec:thermoanalysis}

Einstein-Maxwell theory with a complex scalar field (with a given scalar field mass and electric charge) is described by action \eqref{action}. The only asymptotically flat solutions of this theory are Minkowski spacetime and the Reissner-Nordstr\"om (RN) black hole family of solutions, which have a vanishing scalar field. However, if we introduce a reflecting box or mirror that confines the scalar field, we have found that the theory admits several regular asymptotically flat solutions. These are the caged RN BH (Section \ref{sec:cagedRN}), the ground state soliton (Section \ref{sec:soliton}), the ground state hairy black hole (Section \ref{sec:HairyBH}) and an infinite tower of excited solitons and hairy black holes\footnote{Recall that the ground state solutions have their perturbative root in the lowest normal mode frequency of Minkowski spacetime, and the excited states emerge from the remaining infinite tower of normal mode frequencies.}. For a given scalar field electric charge, the latter excited solutions always have larger energy than the ground state solutions so we do not discuss them further.  

It is natural to expect that the thermal phases of the theory compete with each other. It is thus important to display the several solutions of the theory in a phase diagram of (regular) asymptotically flat boxed solutions. For example, it is important to display the region of existence of each of these solutions in a mass-charge phase diagram as well as to present a microcanonical phase diagram whereby we plot the entropy of the solutions as a function of their mass and electric charge.\footnote{RN black holes caged inside a box were discussed in the grand-canonical ensemble in \cite{Braden:1990hw}. Moreover, \cite{Basu:2016srp} constructed numerically the hairy solutions of the system and discussed them in the  grand-canonical ensemble. At least for small charges where our perturbative analysis is valid, hairy black holes can be the dominant thermal phase only in the microcanonical ensemble so we restrict our discussion to this ensemble.} 

For this end, we find appropriate to work with the Brown-York quasilocal mass $\mathcal{M}$ and charge $\mathcal{Q}$. Essentially, this is because these quasilocal quantities are independent of the parameter $\eta$ that describes the energy-momentum content of the shell that confines the scalar field.\footnote{Essentially, these different solutions differ from each other in the interior region and the energy-momentum content of this region is best captured by the quasilocal thermodynamics. Further recall that these quasilocal quantities  obey the quasilocal first law of thermodynamics \eqref{1stlawBH} and \eqref{1stlawsoliton}, which we used to check our final results.} The quasilocal thermodynamics of the (ground state) soliton, caged RN black hole,  and hairy black hole are given in \eqref{ThermoSoliton}, \eqref{RNQLThermo} and \eqref{ThermoHairyBH}, respectively (for vanishing scalar field mass $m=0$).

\begin{figure}[th]
\centerline{
\includegraphics[width=.48\textwidth]{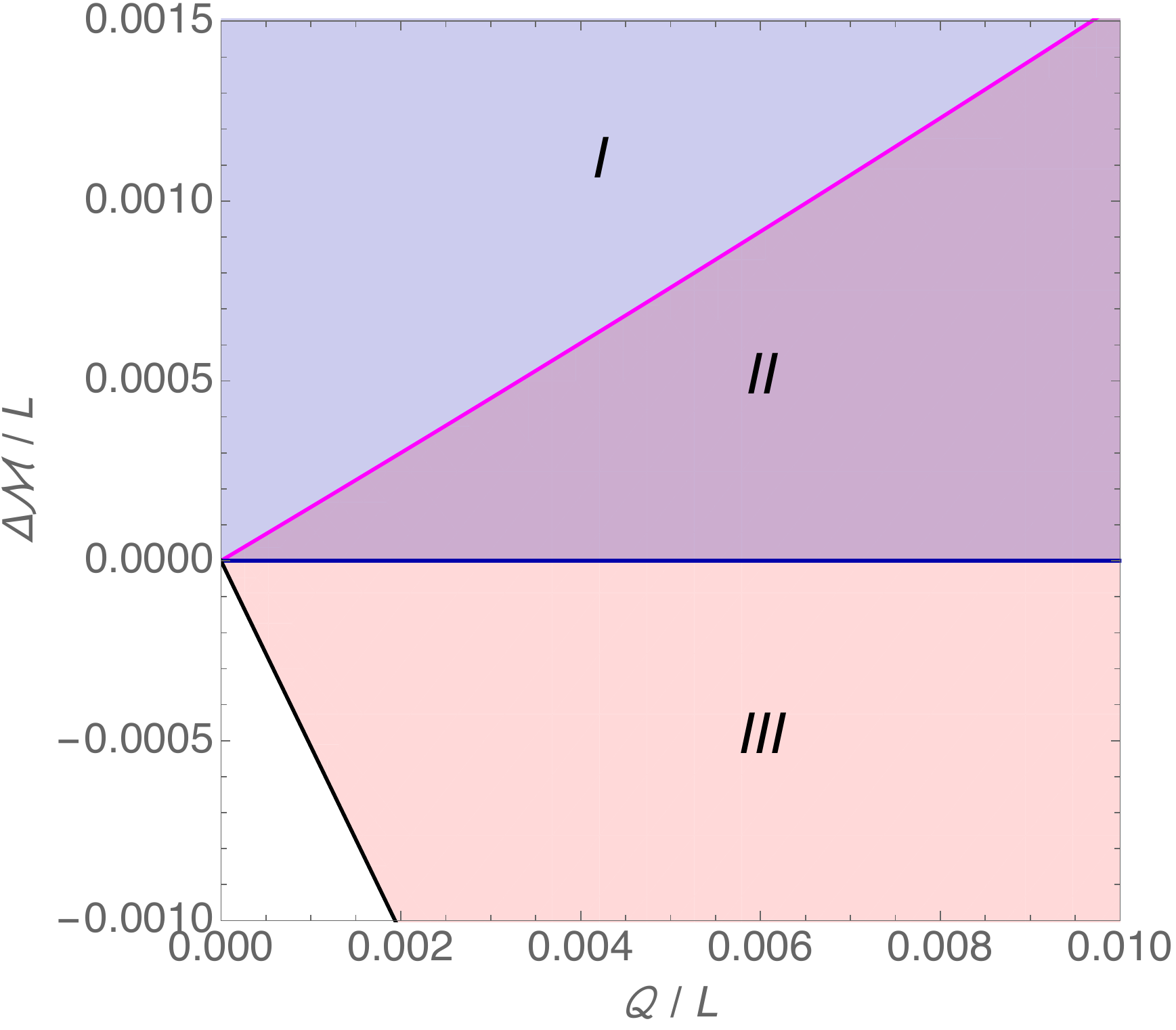}
\hspace{0.3cm}
\includegraphics[width=.48\textwidth]{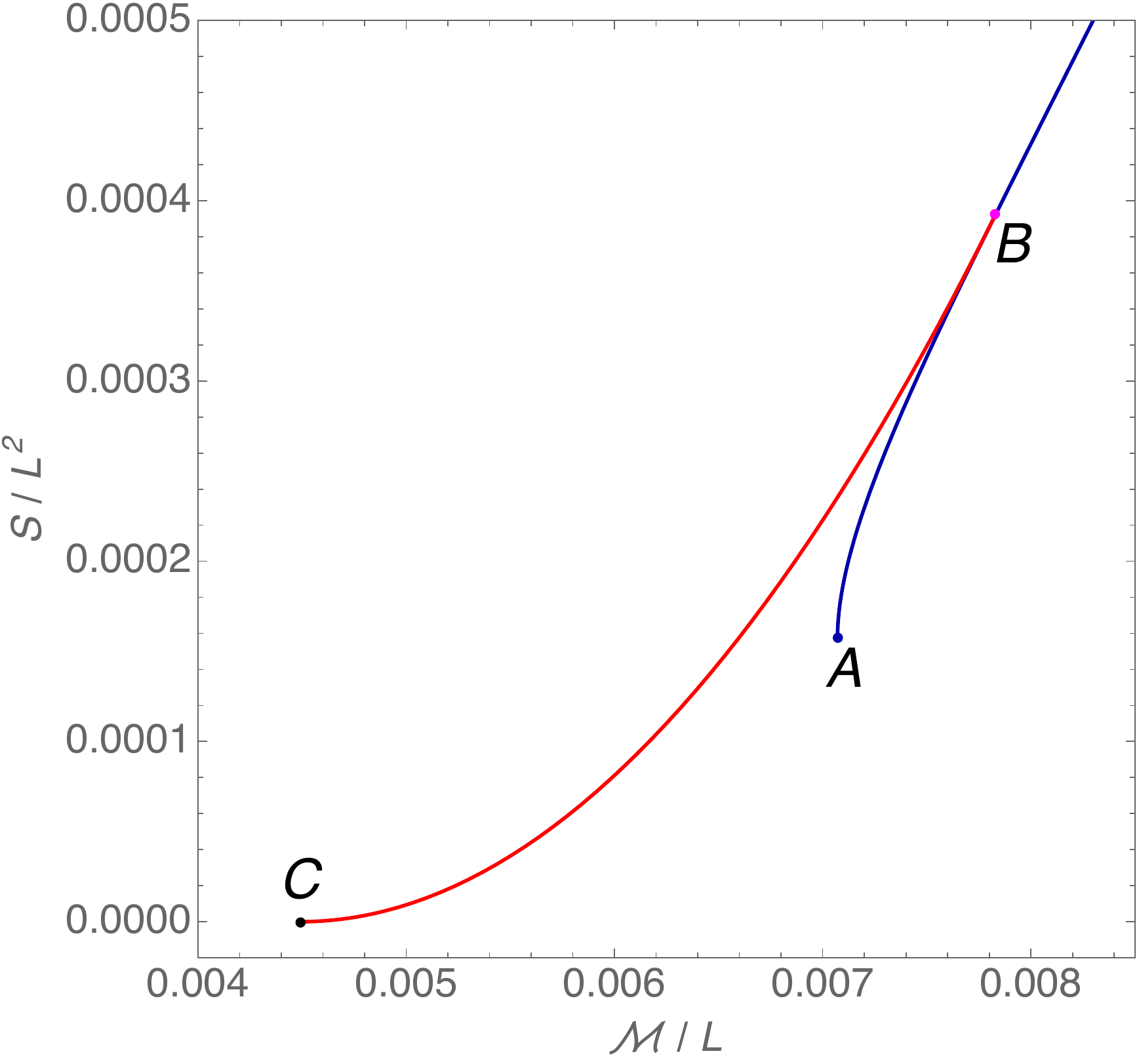}
}
\caption{{\bf Left Panel:} Region of existence of solutions. We plot the quasilocal mass difference  $\Delta {\cal M=M-M}_{ext}$ (between a given solution and the extremal caged RN BH with the same charge $\cal Q$) as a function of the quasilocal electric $\cal Q$  for $e=qL=3.5$. Caged RN black holes exist in regions $I$ and $II$. The magenta line with positive slope splitting these two regions describes the onset curve of the superradiant instability (RN BHs are unstable in region II). The black line with negative slope describes the soliton family. Hairy black holes exist in between these two lines, \ie in regions $II$ and $III$ (red shaded area).  {\bf Right Panel:}  Phase diagram in the microcanonical ensemble \ie  adimensional entropy $S/L^2$ as a function of the adimensional quasilocal mass ${\cal M}/L$ at constant value of the quasilocal charge ${\cal Q}/L=0.005$ and $e=3.5$. The blue line starting at $A$ (extremality) is the RN black hole and the red line $BC$ is the hairy back hole branch that ends at the soliton $C$. The merger point $B$ (superradiant instability onset) signals a second order phase transition. }
\label{fig:MCphaseDiag}
\end{figure}

The left panel of Fig. \ref{fig:MCphaseDiag} shows the region of existence of caged RN black holes, (ground state) solitons and (ground state) hairy BHs for a scalar field charge $e\equiv q L=3.5$. The diagram is qualitatively similar for other values of $e$. The horizontal axis  scans the dimensionless quasilocal charge $\mathcal{Q}/L$ while the vertical axis plots the difference $\Delta {\cal M\equiv M-M}_{ext}$ between the dimensionless quasilocal mass $\mathcal{M}/L$ of a given solution and the mass ${\cal M}_{ext}$ of the extremal caged RN BH that has the same electric charge. So, the extremal caged RN BH family is represented by the horizontal blue line with $\Delta {\cal M}=0$. 
RN BHs exist above the blue line, in regions $I$ and $II$. In region II, RN BHs are unstable, with the magenta line separating regions I and II describing the onset of superradiance. The black line (with negative slope) describes the hairy soliton branch. Hairy black holes exist in regions II and III. That is, they merge with the RN BH family at the onset of superradiance $-$ described by \eqref{ThermoHairyBH} with $\varepsilon=0$ $-$ and extend all the way down to the soliton line $-$ described by \eqref{ThermoHairyBH} with $R_+=0$. In region $II$ we have non-uniqueness of solutions for a given quasilocal mass and charge. 

In the right panel of Fig. \ref{fig:MCphaseDiag} we plot the dimensionless entropy $S/L^2$ as a function of the quasilocal mass ${\cal M}/L$ for a fixed quasilocal charge ${\cal Q}/L=0.005$, again for $q L=3.5$. This phase diagram is qualitatively similar for other (small values) ${\cal Q}/L$ and $q L$. The caged RN BH branch (blue curve) exists for large mass and charge and extends all the way down to point $A$. This point $A$ represents the extremal RN configuration with $T=0$ but finite entropy (it corresponds to the point with ${\cal Q}/L=0.005$ in the blue horizontal line of the left panel). Point $B$ signals the onset of superradiance: caged RN BHs between point $B$ and $A$ are unstable. The onset of superradiance (point $B$) signals a second order phase transition to the branch of hairy black holes (which is stable to superradiance) and extends from point $B$ towards point $C$ (that has vanishing entropy). The latter, is the limit $R_+\to 0 $ of the hairy BH family and describes the hairy soliton with ${\cal Q}/L=0.005$. The former (point $B$) describes the hairy BH branch in the limit $\varepsilon \to 0$ where it merges with the RN family. 

In the microcanonical ensemble the  energy $\cal M$ and the charge $\cal Q$  are held fixed and the relevant thermodynamic potential is the entropy $S$. The thermal phase that has the highest entropy for a given mass and charge is the dominant phase in this ensemble (recall that the soliton and Minkowski spacetime in a box have vanishing entropy as they do not have a horizon). Therefore, from the right panel of  Fig. \ref{fig:MCphaseDiag} we conclude that hairy BHs are the favoured thermal phase in the microcanonical ensemble. In particular, this is true in the region of phase space where they coexist with caged RN black holes. This dominance extends to all other values of the electric charge ${\cal Q}/L$ (where our perturbative analysis holds) and $e=q L$. Thus hairy black holes are the dominant phase in the microcanonical ensemble in their region of existence (regions II and III in the left panel of Fig. \ref{fig:MCphaseDiag}). These hairy black holes are stable to superradiance and we do not have any arguments suggesting they are unstable to any other mechanism within the theory described by the action \eqref{action}. 

The regime of validity of our perturbative computations can be inferred from the right panel plot of Fig. \ref{fig:MCphaseDiag}. Indeed, recall that the (quasilocal) first law of thermodynamics requires that at a second order phase transition the slope $\mathrm{d}S/\mathrm{d}{\cal M}$ is the same for the two branches of solutions that merge (since $T$ is the same at the bifurcation point and $\mathrm{d}{\cal Q}=0$ in the right panel of Fig. \ref{fig:MCphaseDiag}). This is clearly the case in our plot. However, if we start departing from the regime of validity of our perturbative analysis we increasingly find that the merger is not perfect and the slopes of the two branches no longer match. We find that this is the best criteria to identify the regime of validity of our thermodynamic results \eqref{ThermoHairyBH}. With this criteria we can quantize the regime of validity. Typically, by construction, our perturbative results are valid for $\varepsilon\ll 1$ and $R_+\ll 1$. Moreover, the scalar field charge cannot be too large to avoid large back reactions. Altogether, we restrict ourselves to $\varepsilon\lesssim 0.1$, $R_+\lesssim 0.1$ and $\frac{\pi}{\sqrt 2}\leq e\lesssim 4$. With these bounds into consideration we find that for a scalar field charge $e=3.5$ the mass and the charge must be below $\lesssim 0.07$ and $\lesssim 0.05$, respectively.

Our findings allow for a solid expectation about the endpoint of the superradiant instability of RN BHs caged in a box in asymptotically flat backgrounds. Consider the time evolution of a caged RN BH in the segment $AB$ which is perturbed by a charged scalar field. As caged RN BHs in this region are unstable to superradiance, they should  evolve to another black hole solution that is stable to this mechanism. The second law of thermodynamics implies that the entropy can only increase. Then we should expect the unstable RN BH to evolve to the hairy black hole solution we have constructed. 

%%%%%%%%%%%%%%%%%%%%%%%%%%%%%%
\section{Conclusion and final discussions}\label{sec:conc}

Reissner-Nordstr\"om black holes placed inside a cavity are unstable if they are perturbed by a charged scalar field. We have considered the case where the scalar field is completely confined inside the box, {\it i.e.} when it vanishes at and outside the mirror, following the original spirit of the black hole bomb setup proposal. In a phase diagram of asymptotically flat solutions, we have shown that the onset of the instability signals a bifurcation to a new family of solutions that describes hairy black holes. These are asymptotically flat and regular everywhere in the outer domain of communications and they are stable to the mechanism that drives the original Reissner-Nordstr\"om unstable. Therefore, the box and its confinement boundary conditions for the scalar field allow to evade the original no-hair theorems for the Einstein-Maxwell-scalar theory \cite{Ruffini:1971bza,Chrusciel:1994sn,Bekenstein:1996pn,Heusler:1998ua}. That is to say, 
Reissner-Nordstr\"om black holes are {\it not} the only asymptotically flat, spherically symmetric and static (regular) black hole solutions of Einstein-Maxwell theory. The zero horizon radius limit of our hairy black holes is a regular asymptotically flat horizonless solution, {\it i.e.} a hairy soliton (aka boson star after a $U(1)$ gauge transformation).
To be physically relevant, these hairy solutions must obey the energy conditions. This constrains the energy-momentum tensor of the box that we use to confine the hair. We found the Israel stress surface tensor of this box and the conditions it must satisfy to obey the energy conditions. As a byproduct of our study, we found that these hairy black holes must obey a Brown-York quasilocal version of the first law of thermodynamics. 

We found the hairy black holes of the theory for small mass and electric charge. 
It would be interesting to find these hairy solutions for larger values of mass and electric charge (this necessarily requires a numerical construction). The reason being that, above a critical mass and charge, it might well be the case that the zero horizon radius limit of the hairy black hole solutions is singular and no longer a soliton (this scenario occurs in the AdS hairy black holes of \cite{Dias:2011tj}).

In the region of phase space where Reissner-Nordstr\"om black holes are unstable, there is a hairy black hole (with same energy and charge) that always has higher entropy. Therefore, in agreement with the second law of thermodynamics, it is natural to expect that our hairy black holes  are the endpoint of the superradiant instability of  Reissner-Nordstr\"om black holes confined in a box. Thus, in the future it would be important to compare late time solutions of Cauchy simulations like those of \cite{Sanchis-Gual:2015lje,Sanchis-Gual:2016tcm,Sanchis-Gual:2016ros} with our results.
Such time evolution studies might also find useful to monitor the Brown-York quasilocal quantities and first law. Finally, in these time simulations it would be  important  to study carefully  the Israel surface stress tensor of the box that confines the scalar field and the conditions in which the energy conditions are obeyed.

%%%%%%%%%%%%%%%%%%%%%%%%%%%%%%%%%%%%%
%%%%%%%%%%%%%%%%%%%%%%%%%%%%%%%%%%%%%
\vskip .5cm
\centerline{\bf Acknowledgements}
\vskip .2cm
We wish to thank Ian Hawke, Jorge Santos, Kostas Skenderis and Marika Taylor for discussions. %O.J.C.D. and R.M.
The authors acknowledge financial support from the STFC Ernest Rutherford grants ST/K005391/1 and ST/M004147/1. O.D. further acknowledges support from the STFC ``Particle Physics Grants Panel (PPGP) 2016", grant ref. ST/P000711/1. 

%%%%%%%%%%%%%%%%%%%%%%%%%%%%%%%%%%%%%
%%%%%%%%%%%%%%%%%%%%%%%%%%%%%%%%%%%%%

%%%%%%%%%%%%%%%%%%%%%%%%%%%%%%%%%%%%%
%\begin{appendix}

%%%%%%%
%\end{appendix}

%%%%%%%%%%%%%%%%%%%%%%%%%%%%%%%%%%%%%%%%%%
%%%%%%%%%%%%%%%%%%%%%%%%%%%%%%%%%%%%%%%%%%

\bibliography{refs_BoxQN}{}
\bibliographystyle{JHEP}

%%%%%%%%%%%%%%%%%%%%%%%%%%%%%%%%%%%%%%%%%%
%%%%%%%%%%%%%%%%%%%%%%%%%%%%%%%%%%%%%%%%%%
\end{document}